\documentclass[twocolumn,superscriptaddress,showpacs,prd,aps,amsmath,amssymb]{revtex4-1}
\usepackage{natbib}
\usepackage{graphicx,color}
\usepackage{amsmath,amssymb}
\usepackage{verbatim}
\usepackage{float}
\usepackage{wasysym}
\usepackage{amssymb,graphicx}
\usepackage{epsfig}
\usepackage{psfrag}
\usepackage{dsfont}
\usepackage{amsfonts}
\usepackage{mathrsfs}
\usepackage{multirow}
\usepackage{times}
\usepackage{bm}
\usepackage{hyperref}
\hypersetup{
  colorlinks=true,        
  linkcolor=blue,         
  citecolor=cyan,         
}
\begin{document}
\title{Effects of spin on magnetized binary neutron star mergers and jet launching}
\author{Milton Ruiz}
\affiliation{Department of Physics, University of Illinois at
  Urbana-Champaign, Urbana, IL 61801}
\author{Antonios Tsokaros}
\affiliation{Department of Physics, University of Illinois at
  Urbana-Champaign, Urbana, IL 61801}
\author{Vasileios Paschalidis} \affiliation{Departments of Astronomy
  and Physics, University of Arizona, Tucson, AZ 85719}
\author{Stuart L. Shapiro}
\affiliation{Department of Physics, University of Illinois at
  Urbana-Champaign, Urbana, IL 61801}
\affiliation{Department of Astronomy \& NCSA, University of
  Illinois at Urbana-Champaign, Urbana, IL 61801}

%
\begin{abstract}
  Events GW170817 and GRB 170817A provide the best confirmation so far
  that compact binary mergers where at least one of the companions is
  a neutron star can be the progenitors of short gamma-ray bursts
  (sGRBs).
  An open question for GW170817 remains the values and impact of the initial
  neutron star spins. The initial
  spins could possibly  affect
  the remnant black hole mass and spin, the remnant disk and the
  formation and lifetime of a jet and its outgoing electromagnetic
  Poynting luminosity. Here we summarize our general relativistic
  magnetohydrodynamic simulations of spinning, neutron star binaries
  undergoing merger and delayed collapse to a black hole. The binaries
  consist of two identical stars, modeled as $\Gamma=2$ polytropes, in
  quasicircular orbit, each with spins
  $\chi_{\rm{NS}}=-0.053,\,0,\,0.24$, or $0.36$.  The stars are endowed
  initially with a dipolar magnetic field extending from the interior
  into the exterior, as in a radio pulsar. Following merger, the
  redistribution of angular momentum by magnetic braking and magnetic
  turbulent viscosity in the hypermassive neutron star (HMNS) remnant,
  along with the loss of angular momentum due to gravitational
  radiation, induce the formation of a massive, nearly uniformly
  rotating inner core surrounded by a magnetized Keplerian disk-like
  envelope. The HMNS eventually collapses to a black hole, with spin
  $a/M_{\rm BH} \simeq 0.78$ independent of the initial spin of the
  neutron stars, surrounded by a magnetized accretion disk. The larger
  the initial neutron star spin the heavier the disk. After $\Delta
  t\sim 3000M-4000 M \sim 45(M_{\rm NS}/1.625 M_\odot)\rm ms-60(M_{\rm
    NS}/1.625M_\odot)\rm ms$ following merger, a mildly relativistic
  jet is launched.  The lifetime of the jet [$\Delta t\sim 100(M_{\rm
      NS}/1.625M_\odot){\rm ms}-140(M_{\rm NS} /1.625M_\odot)\rm ms$]
  and its outgoing Poynting luminosity [$L_{\rm EM}\sim 10^{51.5\pm
      1}\rm erg/s$] are consistent with typical sGRBs, as well as with
  the Blandford--Znajek mechanism for launching jets and their
  associated Poynting luminosities.
\end{abstract}

\maketitle

\section{Introduction}

The gravitational wave (GW) detection
GW170817~\cite{TheLIGOScientific:2017qsa}~coincident with
electromagnetic (EM) counterpart radiation across the EM spectrum and,
in particular, the detection of the short gamma-ray burst (sGRB) 1.7s
following the inferred merger time by the Fermi Gamma-Ray Burst
Monitor~\cite{FERMI2017GCN} and INTEGRAL~\cite{Savchenko:2017ffs,
  Savchenko17GCN} (event GRB 170817A), provide the best confirmation
so far that compact binary mergers, in which at least one of the
binary companions is a neutron star, can be the progenitors of sGRBs,
as anticipated in~\cite{Pac86ApJ,EiLiPiSc,NaPaPi}. We recently
demonstrated this possibility by self-consistent simulations in full
general relativistic magnetohydrodynamics (GRMHD) of merging black
hole-neutron star (BHNS) binaries~\cite{prs15, Ruiz:2018wah}, and
merging neutron star binaries (NSNS)~\cite{Ruiz:2016rai,Ruiz:2017inq}.
Depending on the spin priors of the binary companions, the GW170817
inferred masses are in the broad range of $0.86M_\odot-2.26M_\odot$,
though the total mass of the system is constrained to be~$2.73
M_\odot-3.29 M_\odot$ with
$90\%$~credibility~\cite{TheLIGOScientific:2017qsa}.  These masses are
consistent with astrophysical observations of NSs~(see
e.g.~\cite{Ozel:2016oaf, Lattimer:2015nhk} and references therein)
which, along with the optical counterparts
\cite{FERMI2017GCN,Savchenko:2017ffs,Savchenko17GCN,GBM:2017lvd},
indicate the presence of matter, and hence strongly suggest the
coalescence of a NSNS as the progenitor of GW170817, although it
cannot rule out the possibility that one of the binary companions is a
stellar-mass BH~(see
e.g.~\cite{Yang:2017gfb,Hinderer:2018pei,Foucart:2018rjc,Coughlin:2019kqf}).

The GRMHD simulations of BHNSs reported in~\cite{prs15,Ruiz:2018wah}, in which the NS is modeled
as an irrotational $\Gamma=2$ polytrope, showed  that an incipient jet --a collimated, mildly
relativistic outflow which is magnetically dominated (i.e. $b^2/(2\rho_0)>1$,
where $b^2=B^2/4\pi$, $\rho_0$ is the rest-mass density, and $B^2=B_i\,B^i$, with $B^i$ the
magnetic field)-- may be launched from the highly spinning BH + disk remnant if: a) the
NS is endowed with a magnetic field that extends from the stellar interior into the exterior,
as  in a radio pulsar; b) the tilt angle between the magnetic moment and the total angular
momentum of the system is small; and c) the initial BH spin satisfies~$a/M_{\rm BH}\gtrsim 0.4$.

Note that the GRMHD simulations in~\cite{prs15,Ruiz:2018wah} do not account for all the physical
processes involved in BHNS mergers, such as a realistic finite--temperature nuclear equation of
state (EOS), neutrino processes, etc.  It has been suggested that neutrino pair annihilation in
BH + disk engines may carry away a significant amount of energy from inner regions of the disk
that may be strong enough to power jets~\cite{Popham:1998ab,Matteo:2002ck,Chen:2006rra,Lei2013ApJ,
  Just:2015dba}, though their typical energies and durations might be too small to explain the
majority of sGRBs~\cite{Just:2015dba}. Recently, it was suggested in
\cite{Lei:2017zro} that the emergence of a jet in a slowly spinning BH + disk engine may
be  dominated initially by neutrino pair annihilation followed by the Blandford--Znajek (BZ)
\cite{BZeffect77} process, leading to a transition from a thermally dominated fireball to a
Poynting dominated outflow as observed in some GRBs such as GRB 160625B~\cite{Dirirsa:2017pgm}. 

On the other hand, the GRMHD studies reported
in~\cite{Ruiz:2016rai,Ruiz:2017inq}, where the NS is modeled as an
irrotational $\Gamma=2$ polytrope, showed that NSNS systems may launch
an incipient jet whether or not the seeded poloidal magnetic field is
confined to the NS interior as long as the binary undergoes delayed
collapse to a BH~\footnote{Note that it has been
  suggested~\cite{Totani:2013lia,Paschalidis:2018tsa} that NSNS
  undergoing prompt collapse to a BH may be the progenitors of fast
  radio bursts, a new class of radio transients lasting between a few
  to a couple of tens of
  milliseconds~\cite{Lorimer:2007qn,Thornton:2013iua}.}.  The lifetime
of the jet [$\Delta t\sim 100(M_{\rm NS}/1.625M_\odot){\rm ms}$] and
the outgoing electromagnetic luminosities~[$L_{\rm EM}\sim 10^{51}\rm
  erg/s$] in the above cases turn out to be consistent with
short-duration sGRBs~\cite{Bhat:2016odd,Lien:2016zny,Svinkin:2016fho}.
Note that the GRMHD simulations reported
in~\cite{Kawamura:2016nmk,Ciolfi:2017uak}, where the effects of
different EOSs, different mass ratios, and different orientations of a
poloidal magnetic field confined to the NS interior were probed, lack
of an outgoing outflow or jet was observed, though formation of an
organized magnetic field structure above the BH was evident (see
e.g.~Fig. 9 in \cite{Kawamura:2016nmk}). It is likely that the lack of
a jet is due to insufficient resolution to properly capture the
magnetic instabilities that boost the magnetic field strength
to~$\gtrsim 10^{15.5}\rm G$, an essential ingredient for jet
launching~\cite{prs15,Ruiz:2016rai}.  On the other hand, the very high
resolution NSNS mergers reported in~\cite{Kiuchi:2014hja}, where NSs
are modeled by an H4 EOS and endowed with a poloidal magnetic field
confined to the NS interior, did not find any evidence of a
magnetically-driven outflow after about $\sim 39\rm ms$ following
merger. The lack of a jet in these studies has been attributed to the
persistent fall-back debris in the atmosphere, which increases the ram
pressure above the BH poles. Therefore, a longer simulation is likely
required for jet launching. Moreover, the emergence of a jet may be
possible only for EOSs for which the matter fall-back timescale is
shorter than the accretion disk lifetime
\cite{Paschalidis:2016agf}. Note that heating induced by neutrino pair
annihilation in NSNS mergers is not efficiently translated into
relativistic outflows and, therefore, neutrinos may not be strong
enough to power jets by themselves~\cite{Just:2015dba,Perego:2017fho}.

Due to the limited sensitivity of the second observing run (O2) of
Advanced LIGO~\cite{TheLIGOScientific:2017qsa, GBM:2017lvd}, and
assuming that the progenitor of GW170817 is the merger of a NSNS
system, there is no current consensus yet whether the GW170817 remnant
is a highly spinning BH + disk or a long-lived supramassive NS
(SNS). Depending on the EOS, NSNS mergers may yield a remnant that can
form a long-lived SNS, a transient differentially rotating
hypermassive NS (HMNS) that can survive for many rotation periods, or
promptly collapse to a
BH~\cite{Shibata:2002jb,STU1,Shibata:2006nm}. It was argued
in~\cite{Margalit:2017dij} that a transient HMNS can produce both
blue and red kilonova ejecta expanding at mildly relativistic
velocities, consistent with observations of
GW170817~\cite{FERMI2017GCN,Savchenko:2017ffs,Savchenko17GCN,GBM:2017lvd}.
This hypothesis is supported by the GRMHD NSNS merger simulations
in~\cite{Ruiz:2016rai,Ruiz:2017inq} and~\cite{Ruiz:2017due} where a magnetized
HMNS remnant undergoing delayed collapse and not a SNS appears to be a
prerequisite for jet launching.
This requirement allows us to impose a bound on the maximum mass of
a spherical NS: ${M_{\rm max}^{\rm sph}} \lesssim 2.74/\beta$
\cite{Ruiz:2017due}. Here
$\beta$ is the ratio of the maximum mass of a uniformly rotating star
over the maximum mass of a nonrotating star.  Causality arguments
allow $\beta$ to be as high as $1.27$~\cite{1987ApJ...314..594F,1997ApJ...488..799K},
while most realistic candidate EOS predict $\beta\sim 1.2$ (which is approximately
EOS independent) yielding ${M_{\rm max}^{\rm sph}}$ in
the range $2.16M_\odot-2.28M_\odot$~\cite{Ruiz:2017due}, consistent with estimates arrived
at from other considerations~\cite{Margalit:2017dij,Shibata:2017xdx,Rezzolla:2017aly}.
By contrast, a broad number of GR
hydrodynamic simulations favoring a long-lived, massive SNS surrounded
by a torus were summarized in~\cite{Shibata:2017xdx} to support their
inferred requirement of a strong neutrino emitter that has a
sufficiently high electron fraction to avoid an enhancement of the
ejecta opacity. Recently, it was shown in~\cite{Piro:2018bpl} that a
long-lived SNS remnant is fully consistent with the multi-wavelength
afterglow data taken by different EM observatories $150$ days after
the GW170817 detection.

The LIGO/Virgo observations of GW170817 practically left the
  pre-merger NS spins unconstrained. These could have a
  strong impact on the remnant disk, the final BH spin, the
  lifetime of the transient HMNS, the amount of
  ejected neutron rich matter that can power kilonovae and synthesize
  heavy elements, as well as the formation and lifetime of a
  magnetically-driven jet and the associated outgoing EM Poynting
  luminosity.  Addressing these issues, GRMHD simulations in the dynamical
  spacetimes of spinning NSNSs are necessary. Understanding the
  aforementioned aspects may explain or give new insight regarding
  sGRB phenomenology and the synergy between EM and GW observations.
  Prior work on spinning NSNSs, but without magnetic fields, has been presented 
  in~\cite{Bernuzzi:2013rza,Tacik:2015tja,PhysRevD.95.044045,Dietrich:2015pxa,Dietrich:2017xqb}
  with constraint satisfying initial data, and
  in~\cite{Kastaun2013,Kastaun2015} with constraint violating initial
  data (see also~\cite{bauswein2015exploring} for work in the
  conformal flatness approximation). Work has also been performed on
  eccentric binaries with spinning NSNSs and constraint
  satisfying initial
  data~\cite{East:2015yea,Paschalidis:2015mla,PEFS2016,East:2016zvv}.

With respect to the spin of the BH which was formed after the collapse of the
merger remnant, Refs. \cite{Kastaun2013,Kastaun2015} found that it increases as
the spin of the NSs in the binary increase. In particular \cite{Kastaun2013}
investigate NSs with a $\Gamma=2$ EOS and spins that range from minus one up to 
$1.2$ times the spin that corresponds to the corotating solution and they find 
a $\sim 10\%$ increase in the BH spin. Similar results were reported in 
\cite{Kastaun2015} with more realistic EOSs. A prompt collapse to a BH is 
possible only if the mass of the NSs in the binary is above a certain 
threshold, which depends on the EOS. This highlights the fact that if the 
total mass of the binary is close to the critical mass for prompt collapse, the 
spin of the NSs can have a strong impact on the dynamics of the merger.
  
Here we initiate new investigations by performing fully
  relativistic GRMHD simulations of {\it magnetized}, and {\it
  spinning} NSNS configurations in a quasicircular orbit that undergo
delayed collapse to BH. The binaries are formed by two identical
spinning NSs modeled by a $\Gamma=2$ polytropic EOS and spins~$
\chi_{\rm NS}\equiv J_{\rm ql}/(M/2)^2=-0.053,\,0,\,0.24,\,0.36$, where
$J_{\rm ql}$ is the quasilocal angular momentum of the NS, and $M$ 
is the Arnowitt-Deser-Misner (ADM) mass of the system \cite{Tsokaros:2018dqs}. 
Denoting by $\Sigma_{\rm cor}$ the circulation that corresponds to the corotating
binary with the same ADM mass and at the same separation, these spins correspond
to circulations of $-0.3 \Sigma_{\rm cor},\ \Sigma_{\rm cor},\ 1.6 \Sigma_{\rm cor}$. 
The stars are initially threaded by a
dipolar magnetic field extending from the stellar interior into the
exterior, as in radio pulsars~\cite{Ruiz:2016rai}. To determine the
impact of the magnetically-driven instabilities on the fate of
spinning NSNS mergers, we also consider unmagnetized evolutions of the
above NSNS configurations.

We find that, following merger, the redistribution of angular momentum
by magnetic braking due to winding and magnetic turbulence driven by
the magnetorotational instability (MRI) in the HMNS remnant, along
with the dissipation of angular momentum due to gravitational
radiation, induce the formation of a massive, nearly uniformly
rotating inner core surrounded by a magnetized, Keplerian, disk-like
envelope (similar behavior has been reported in supermassive stars
modeled by a polytropic EOS with $\Gamma\gtrsim 4/3$ in
\cite{Sun:2018gcl}). In all cases, by~$t-t_{\rm mer}\sim 15(M_{\rm
  NS}/1.625M_\odot){\rm ms}-20(M_{\rm NS}/ 1.625M_\odot)\rm ms$
following merger, the HMNS collapses to a BH. Interestingly, we
  find that the nascent BH spin is $a/M_{\rm BH} \simeq 0.78$
  independent of the initial NS spin. The final BH is
surrounded by an accretion disk whose rest-mass depends strongly on
the initial spin of the NSs. We observe that the larger the initial
spin, the heavier the disk. 
In contrast to the magnetized cases which form a black hole,                     
in the unmagnetized cases,
only the HMNS remnant of the antialigned ($\chi_{\rm NS}=-0.053$) and
the irrotational configurations (those with less centrifugal support)
collapse to a BH during the time evolved 
(in $t-t_{\rm mer}\sim15(M_{\rm NS}/1.625M_\odot){\rm ms}$). 
This is in agreement with the pure hydrodynamic simulations of
\cite{Kastaun2013} whose lowest mass model
($M_{\rm NS}=1.63\ M_\odot$ in their notation, compared to our $1.5\ M_\odot$
model) reaches a quasistationary state, with the BH mass and spin changing less 
than $0.4\%$ during the last $t\sim 50\ M_{\rm BH}$ for all their simulations and 
with negligible mass accretion. 
In our other two cases, the HMNS remnant is driven to a
quasiaxisymmetric configuration on a dynamical timescale and remains
in quasistationary equilibrium until the termination of our
simulations [$t-t_{\rm mer}\gtrsim 60(M_{\rm NS}/1.625M_\odot){\rm
    ms}$]. Angular momentum loss due to gravitational radiation alone
is, therefore, an inefficient mechanism to trigger the collapse of a
highly spinning HMNS. Other dissipation, such as turbulent viscosity or magnetic
fields, will lead to collapse, but only on a longer timescale.

After $\Delta t\sim 3000M-4000M$ $\sim 45(M_{\rm NS}/1.625 M_\odot)\rm
ms-60(M_{\rm NS}/1.625 M_\odot)\rm ms$ following merger, a
magnetically-driven and sustained incipient jet is launched. The
lifetime of the jet [$\Delta t\sim 100(M_{\rm NS}/1.625M_\odot){\rm
    ms}-140(M_{\rm NS}/1.625M_\odot)\rm ms$] and its respective
outgoing EM Poynting luminosity [$L_{\rm EM}\sim 10^{51.5\pm 1}\rm
  erg/s$] turn out to be consistent with typical short-duration sGRBs
(see e.g.~\cite{Bhat:2016odd,Lien:2016zny,Svinkin:2016fho}), as well
as with the BZ process for launching jets and their associated
Poynting luminosities. We also find that the ejecta in the high spin
NSNS configurations (aligned cases) is~$\sim 10^{-2.2}\,M_\odot$ and,
therefore, can give rise to the so-called kilonova event that can be
detected by current telescopes, as well as the Large Synoptic Survey
Telescope (LSST)~\cite{Shibata:2017xdx,Rosswog}.

The remainder of the paper is organized as follows. A short summary of the numerical
methods and their implementation is presented in Sec.~\ref{sec:Methods}. A detailed
description of the adopted initial data and the grid structure  used to solve
the GRMHD equations are given in Sec.~\ref{subsec:idata} and Sec.~\ref{subsec:grid},
respectively.  In Sec.~\ref{subsec:diagnostics}~we describe the diagnostics employed to
monitor and verify the reliability of our numerical calculations. We present our results
in Sec.~\ref{sec:results}. Finally, we summarize our findings and conclusions in Sec.
\ref{sec:conclusion}. Throughout the paper we adopt geometrized units ($G=c=1$)  except where
stated explicitly. Greek indices denote all four spacetime dimensions,
while latin indices imply spatial parts only.
%
%
\begin{center}
  \begin{table}
    \caption{Summary of the initial properties of the NSNS configurations.
      The binaries have ADM mass $M$, ADM angular momentum $J$, while the NSs
      have quasilocal dimensionless spin parameter $\chi_{\rm NS}\equiv J_{\rm ql}/(M/2)^2$
      which is either aligned or antialigned with respect to the total angular
      momentum of the system, approximate rotational period $T$ in units of  $(M_{\rm NS}/1.625
      M_\odot)\rm ms$~\cite{Tsokaros:2018dqs},
      coordinate equatorial radius toward companion $R_x$,
      and coordinate polar radius $R_z$ in units of  $(M_{\rm NS}/1.625 M_\odot)\rm km$.
      In all cases, the orbital separation is fixed at $D_0 = 45 (M_{\rm NS}/1.625 M_\odot)\rm km$,
      corresponding to an initial angular velocity of $M\,\Omega_0 \simeq 0.027$.
      The tag for each configuration is formed by the spin direction (sp = aligned and sm = antialigned)
      followed by its magnitude. For comparison purposes, we also consider the irrotational P--configuration
      treated previously in~\cite{Ruiz:2016rai}, and denoted here as irrot.  
      \label{table:NSNS_ID}}
    \begin{tabular}{cccccc}
      \hline\hline
          Model         & $J/M^2$   &$\chi_{\rm NS}$   & $T\,[\rm ms]$ &$R_x\,[\rm km]$ & $R_z\,[\rm km]$\\  
          \hline
          sp0.36        &  1.14           & $0.36$   &$2.3$   & 15.00 & 13.77 \\
          sp0.24        &  1.09           & $0.24$   &$3.2$   & 14.19 & 13.18 \\          
          sm0.05        &  0.95           & $-0.05$  &$-12.0$  & 13.67 & 12.63  \\
          \hline
          irrot         &  0.98           & $0.0$    & $0.0$   & 13.67 & 12.73  \\
          \hline\hline
    \end{tabular}
  \end{table}
\end{center}
%
\section{Methods}
\label{sec:Methods}
We use the extensively tested Illinois GRMHD code which is
embedded in the \texttt{Cactus}
infrastructure~\cite{AllAngFos01,cactusweb}. The code evolves the
Baumgarte--Shapiro--Shibata--Nakamura (BSSN)
equations~\cite{shibnak95,BS} (for a detailed discussion see
also~\cite{BSBook}) with fourth order centered spatial differencing,
except on shift advection terms, where a fourth order upwind
differencing is used. Outgoing wave-like boundary conditions are
applied to all BSSN evolved variables.  These variables are evolved
using the equations of motion (9)-(13) in~\cite{Etienne:2007jg}, along
with the $1+$log time slicing for the lapse $\alpha$ and the
``Gamma--freezing" condition for the shift $\beta^i$, cast in first
order form (see~Eq.~(2)-(4) in~\cite{Etienne:2007jg}). For numerical
stability, we set the damping parameter $\eta$ appearing in the shift
condition to $\eta=3.75/M$, with $M$ the ADM mass of the system. For
additional stability we modify the equation of motion of the conformal
factor $\phi$ by adding a dissipation term (see Eq.~19 in~\cite{DMSB})
which damps the Hamiltonian constraint. During the whole evolution we
set the constraint damping parameter to $c_H=0.04$.  The time
integration is performed via the method of lines using a fourth-order
accurate Runge-Kutta integration scheme with a Courant-Friedrichs-Lewy
(CFL) factor set to $0.5$.  We use the Carpet
infrastructure~\cite{Carpet,carpetweb} to implement the moving-box
adaptive mesh refinement.  Fifth order Kreiss-Oliger
dissipation~\cite{goddard06} has been also added in the BSSN evolution
equations.

For matter and magnetic field evolution, the code solves the equations
of ideal GRMHD in a conservative scheme via high-resolution shock
capturing methods.  The conservative variables are evolved through
Eqs.  (27)-(29) in~\cite{Etienne:2010ui}. To ensure the magnetic field
remains divergenceless during the evolution, we integrate the magnetic
induction equation using a vector potential $\mathcal{A}^\mu$ (see
Eqs. (19)-(20) in~\cite{Etienne:2010ui}). We adopt the generalized
Lorenz gauge described in~\cite{Farris:2012ux} to avoid the appearance
of spurious magnetic fields~\cite{Etienne:2011re}. The damping
parameter is set to $\xi\approx 6.5/M$. We employ a $\Gamma$--law EOS
$P=(\Gamma-1) \rho_0\,\epsilon$ and allow for shock heating. Here
$\epsilon$ is the specific internal energy and $\rho_0$ is the
rest-mass density.  In all our models we set~$\Gamma=2$.
%
\subsection{Initial data}
\label{subsec:idata}
We consider unmagnetized and magnetized NSNS configurations  in a quasiequilibrium
circular orbit that inspiral, merge and undergo delayed collapse to a BH. The binaries
consist of two identical NSs modeled by a polytropic EOS with $\Gamma=2$. Each binary companion
has an initial spin of $\chi_{\rm NS}=-0.05$, $0.24$, or $0.36$. Our extreme case corresponds to
a binary in which the NSs have an initial rotation period of $\sim 2.3(M_{\rm NS}/1.625 M_\odot)
\rm ms$~(see Table~\ref{table:NSNS_ID}).

The initial data are computed using the Compact Object
CALculator~({\tt COCAL})~\cite{Tsokaros:2015fea,Tsokaros:2018dqs},
and their main properties are listed in Table~\ref{table:NSNS_ID}. 
Following~\cite{Ruiz:2016rai}, we rescale the rest mass of the stars to
$M_{\rm NS}=1.625M_\odot (k/k_L)^{1/2}$ where $k_L=269.6\rm km^2$ is
the polytropic constant used to compute the initial data, and $k=P/\rho_0^\Gamma$. 
In all cases the binaries have ADM mass
$M=4.43(M_{\rm NS}/1.625M_\odot)\rm km$,
and an initial coordinate separation of $45\,(M_{\rm NS}/1.625 M_\odot)\rm km$.  
A single isolated spherical star
with mass $M_{\rm TOV}=M/2$ has compactness ${\cal C}=0.138$, 
second Love number $k_2=0.0807$ \cite{Hinderer:2007mb,Hinderer:2009ca}, 
and tidal deformability $\Lambda= (2/3)k_2{\cal C}^{-5}=1080$. Notice that for 
this EOS with $\Gamma=2$ the maximum mass configuration has a
$\mathcal{C}=0.21$, and $M_{\rm NS}^{\rm max}=1.23 M_{\rm NS}$.
The tag in Table~\ref{table:NSNS_ID} for
each configuration is formed by the spin direction (sp = aligned and
sm = antialigned) followed by its magnitude. For comparison purposes,
we also consider the NSNS P-case treated previously
in~\cite{Ruiz:2016rai}, and denoted~here as irrot. To distinguish
between hydrodynamic or magnetized evolutions, an ``H'' or ``M'' will
precede the tag, respectively.

In the magnetized cases, the stars are initially seeded with a dynamically unimportant  dipole-like
magnetic field generated by the vector potential~\cite{Farris:2012ux,Paschalidis:2013jsa}
\begin{eqnarray}
 A_\phi&=& \frac{\pi\,\varpi^2\,I_0\,r_0^2}{(r_0^2+r^2)^{3/2}}
 \left[1+\frac{15\,r_0^2\,(r_0^2+\varpi^2)}{8\,(r_0^2+r^2)^2}\right]\,,
 \label{eq:Aphi_pot}
\end{eqnarray}
that approximately corresponds to that induced by an interior current
loop with radius $r_0$ and current $I_0$. Here $r^2=\varpi^2+z^2$,
with $\varpi^2= (x-x_{\rm NS})^2 +(y-y_{\rm NS})^2$, and $(x_{\rm NS},
y_{\rm NS})$ is the position of the maximum value of the rest-mass
density of the NS. As in~\cite{Ruiz:2016rai}, we choose $I_0$ and
$r_0$ such that the maximum value of the magnetic-to-gas-pressure
ratio in the NS interior is $\beta^{-1}\equiv P_{\rm mag}/ P_{\rm
  gas}=0.003125$. The initial magnetic field strength at the NS pole
turns out to be ${B}_{\rm pole}\sim
10^{15.2}(1.625M_\odot/M_{\rm NS})\rm G$. This strength has been
chosen in~\cite{Ruiz:2016rai} to mimic the growth of the magnetic
field due to the Kelvin-Helmholtz (KH) instability and the MRI triggered during
the NSNS merger and HMNS phase of the evolution recently reported in
the very high resolution NSNS simulations in~\citep{Kiuchi:2015sga}.
There it was found that during merger the rms magnetic field strength
is boosted from from $\sim 10^{13}\rm G$ to $\sim 10^{15.5}\rm G$,
with local values up to $\sim 10^{17}\rm G$.
%
\begin{figure}
  \centering
  \includegraphics[width=0.46\textwidth]{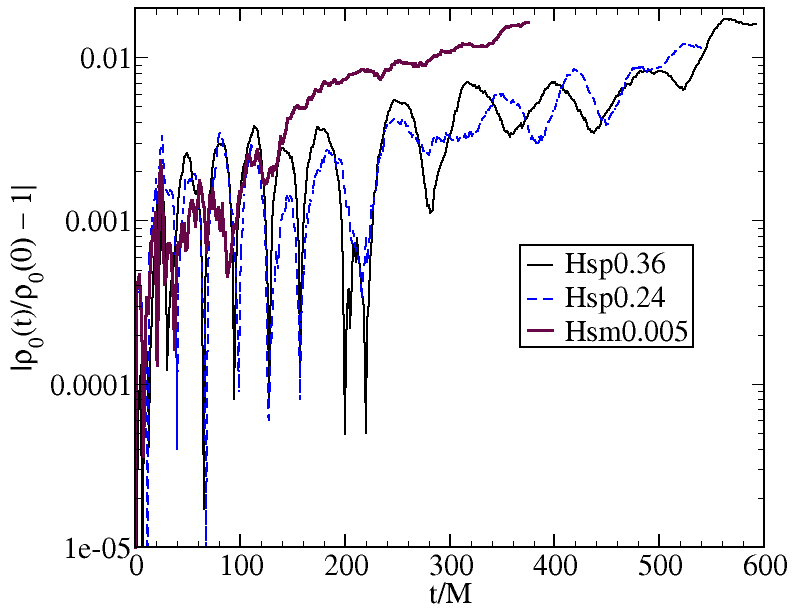}
  \caption{Maximum value of the rest-mass $\rho_0(t)$ during the early inspiral, normalized by the initial
    maximum density $\rho_0(0)$, for the unmagnetized cases (see Table~\ref{table:NSNS_ID}).
    In all cases, the NS oscillations are $\lesssim 1\%$, and more pronounced in the antialigned case
    (Hsm0.005 configuration).
    \label{fig:rho_max_oscillations}}
\end{figure}
%
\begin{table*}
  \caption{Grid hierarchy  in units of $M$ for  models listed in Table~\ref{table:NSNS_ID}. The computational
    mesh consists of two sets of seven nested refinement boxes, the innermost ones  centered on each star.
    The finest box around the NS has a half length of $\sim 1.3\,R_{\rm NS}$, where $R_{\rm NS}$ is the initial
    stellar radius. The number of grid points covering the equatorial radius of NS is denoted by
    $N_{\rm NS}$. In terms of grid points per NS radius the resolution used here is slightly larger than
    that in~\cite{Ruiz:2016rai}. In all cases, we impose symmetry about the orbital plane.}
\begin{tabular}{ccccc}
  \hline \hline
  Model &  Grid Hierarchy (half length)   & Max. resolution  & $N_{\rm NS}$ \\
  \hline \hline
  sp0.36           &  (266.74, 133.37, 66.68, 33.34, 16.67, 8.33, 4.17)     & $0.05M$  &  66\\
  sp0.24           &  (252.34, 126.17, 63.08, 31.34, 15.77, 7.88, 3.94)     & $0.05M$  &  66\\
  sm0.05           &  (243.10, 121.54, 60.77, 30.38, 15.19, 7.60, 3.80)     & $0.05M$  &  66\\
  \hline
  \hline
  irrot$^{(a)}$    &  (246.15, 123.76, 61.53, 30.77, 15.38, 7.70, 3.84)     & $0.05M$  &  61\\
  \hline \hline
\end{tabular}
\begin{flushleft}
  $^{(a)}$ P-case configuration treated previously~in~\cite{Ruiz:2016rai}.
\end{flushleft}
\label{table:grid}
\end{table*}
%

\begin{figure*}
  \centering
  \includegraphics[width=0.33\textwidth]{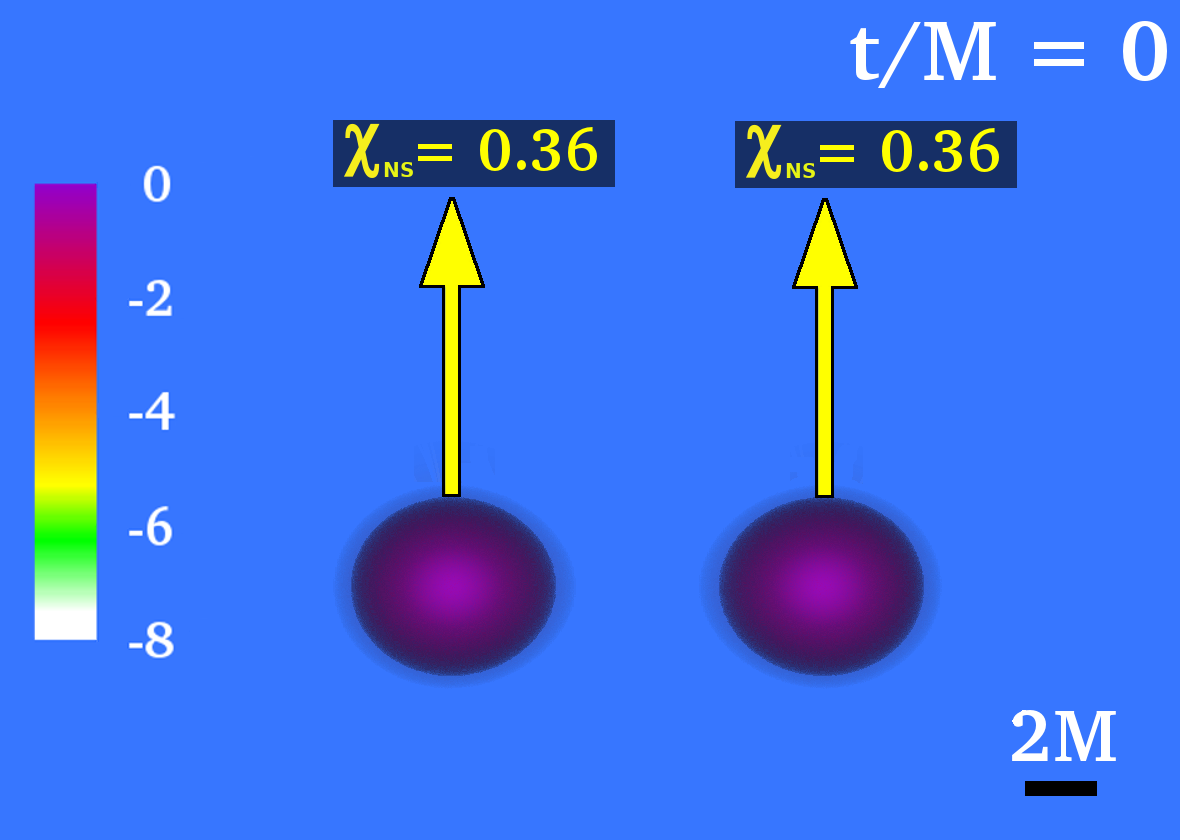}
  \includegraphics[width=0.33\textwidth]{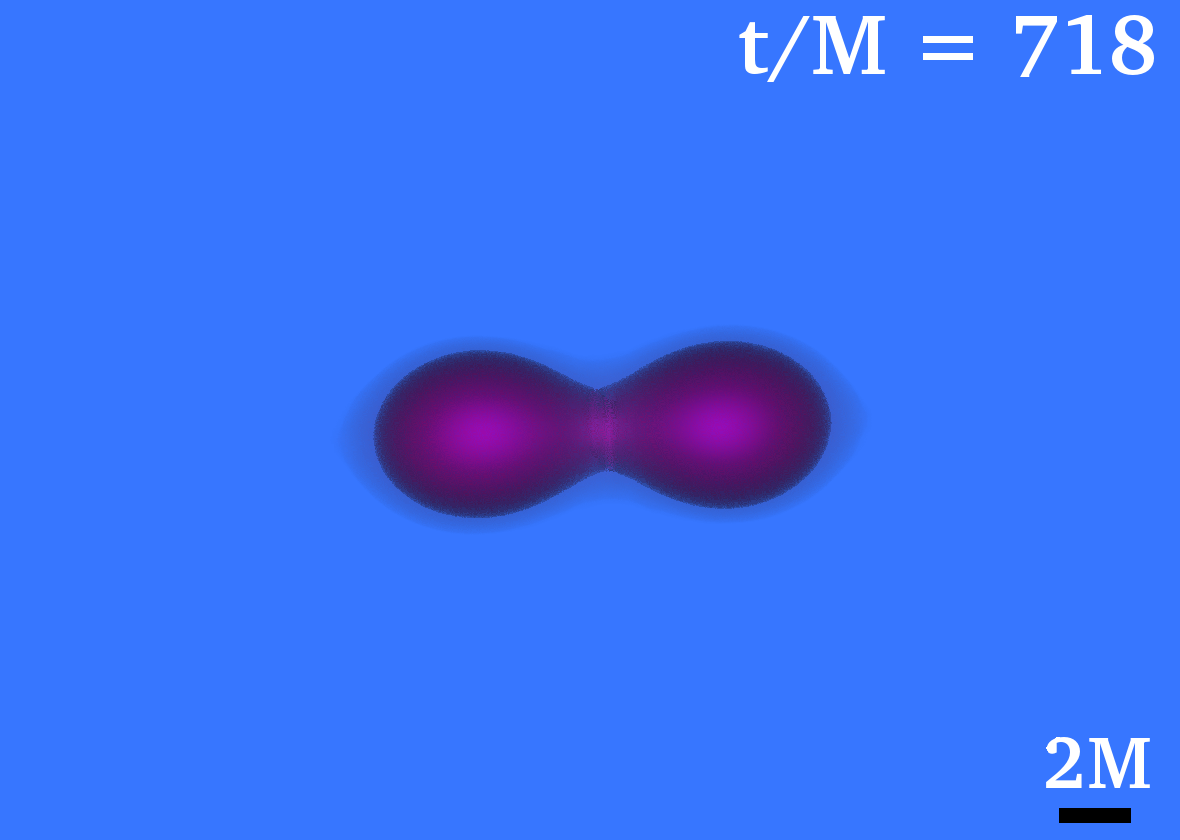}
  \includegraphics[width=0.33\textwidth]{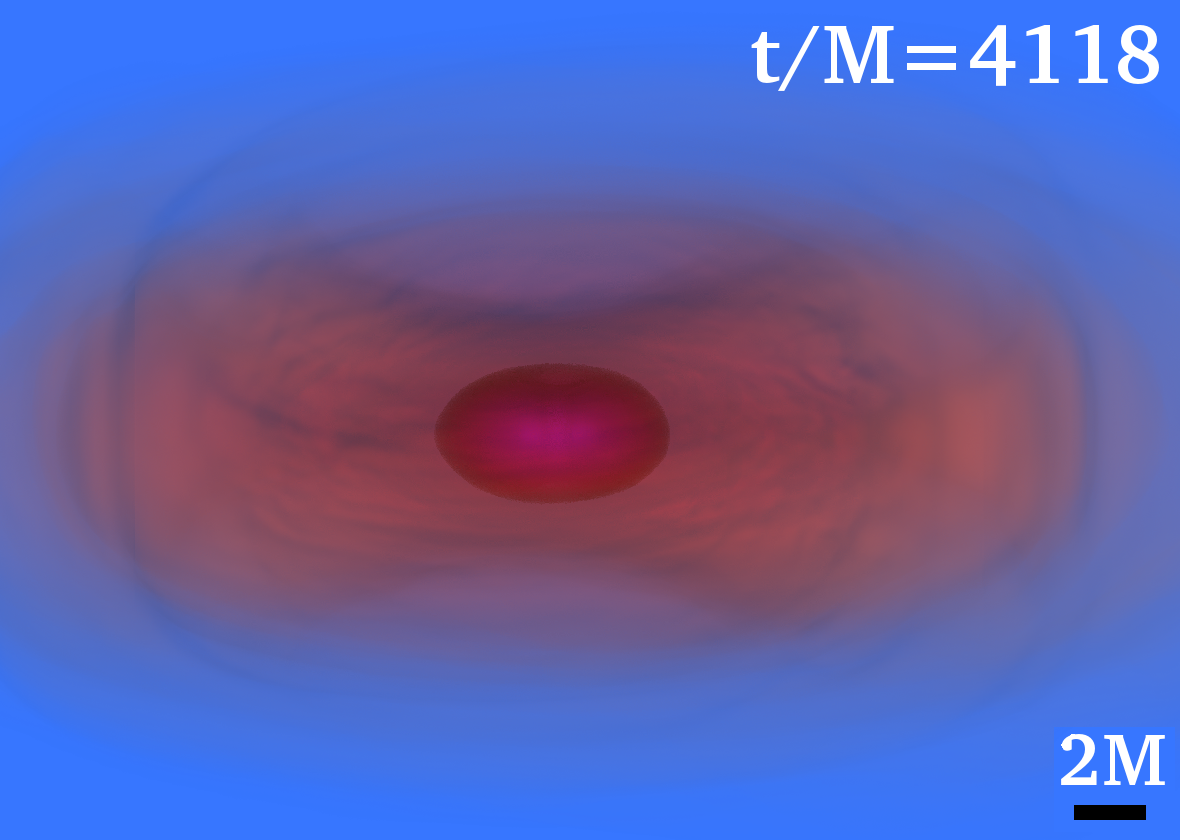}
  \includegraphics[width=0.33\textwidth]{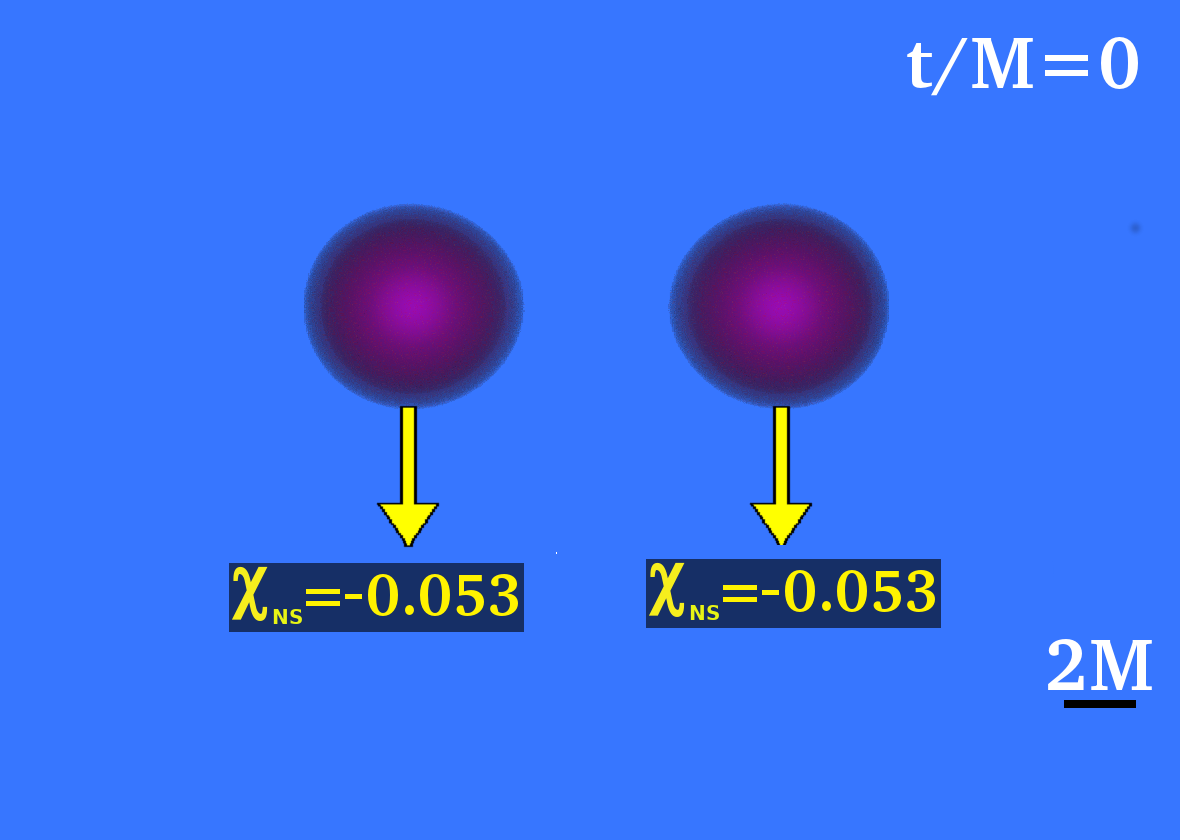}
  \includegraphics[width=0.33\textwidth]{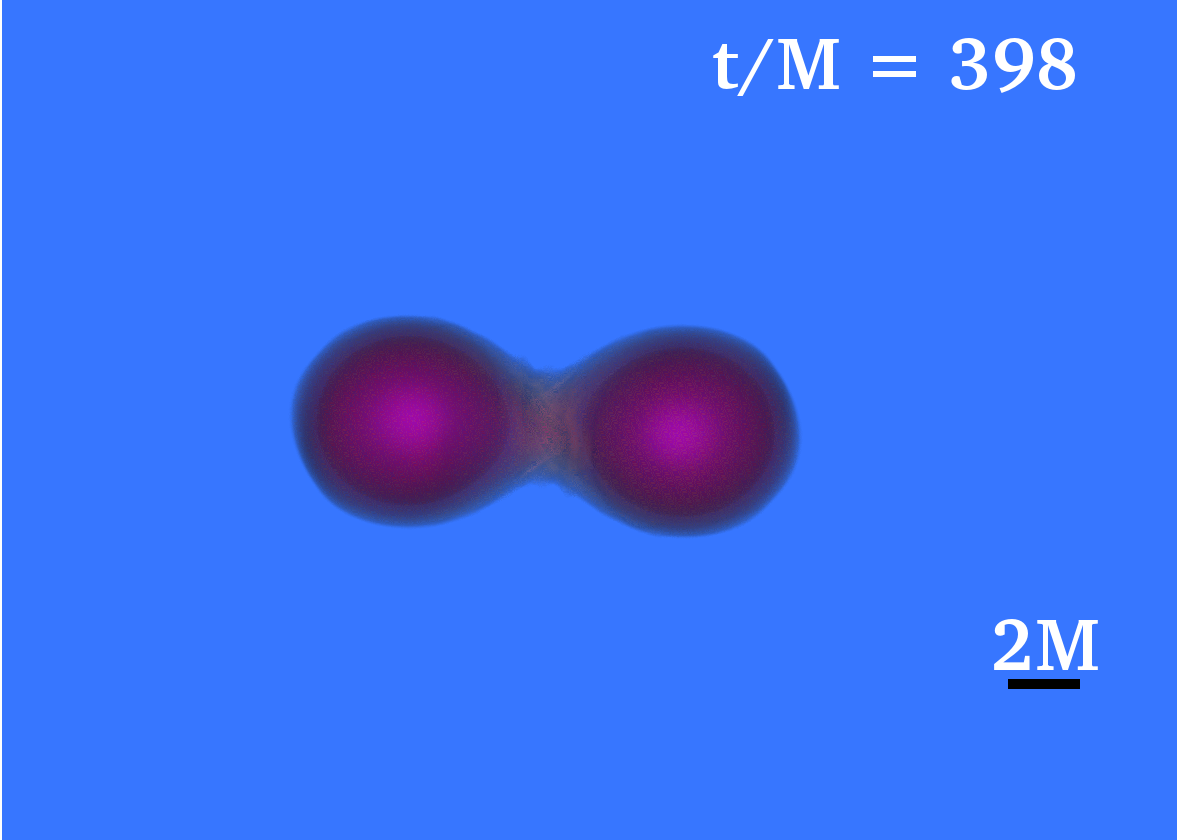}
  \includegraphics[width=0.33\textwidth]{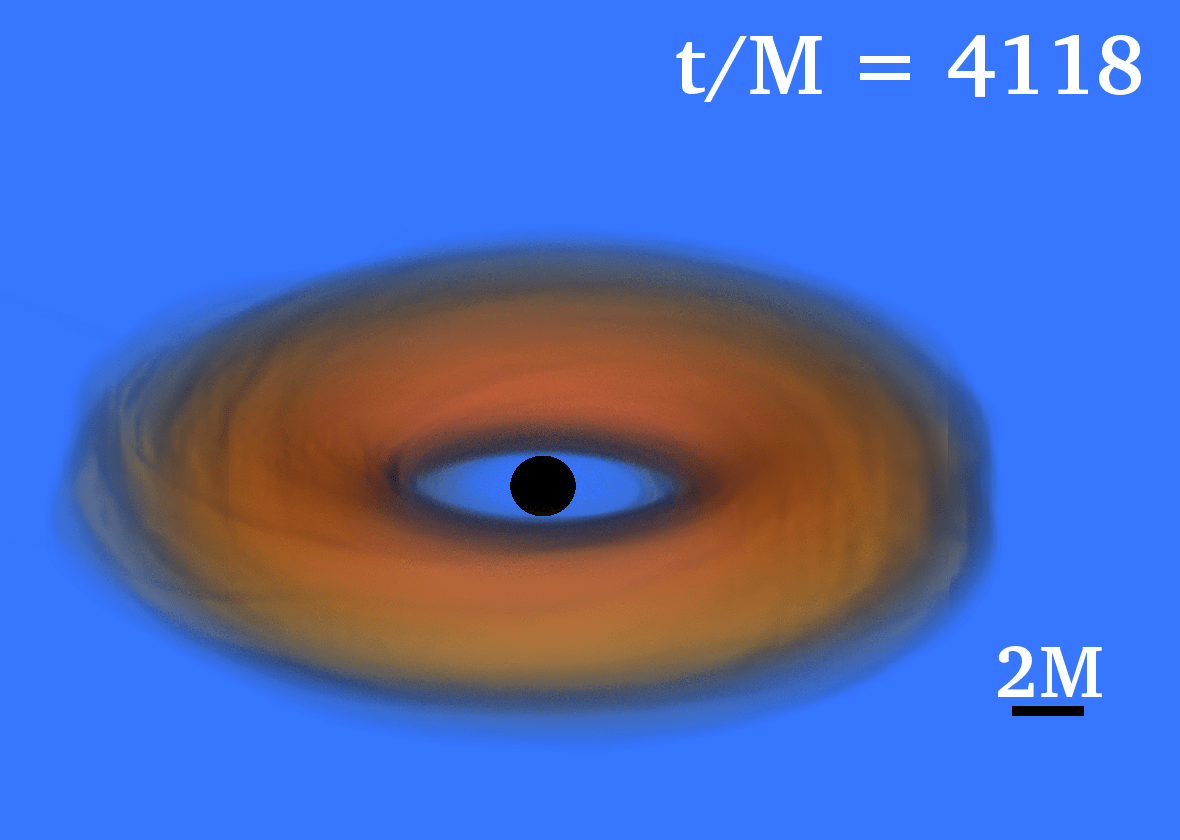}
  \caption{Volume rendering of rest-mass density $\rho_0$, normalized to the initial maximum
    value $\rho_{0,\text{max}}\simeq 10^{14.4}(1.625\,M_\odot/M_{\rm NS})^2\text{g/cm}^3$
    (log scale), at selected times for our extreme unmagnetized cases: Hsp0.36 (top panels)
    and Hsm0.05 (bottom panels). Arrows indicate  the direction of the spin. The BH apparent
    horizon in Hsm0.05 is displayed as a black sphere. Here $M=1.47\times 10^{-2}(M_{\rm NS}/
    1.625M_\odot)\rm ms$ = $4.43(M_{\rm NS}/1.625M_\odot)\rm km$.
    \label{fig:NSNS_snap_hydro}}
\end{figure*}
%
\begin{figure*}
  \centering
  \includegraphics[width=0.33\textwidth]{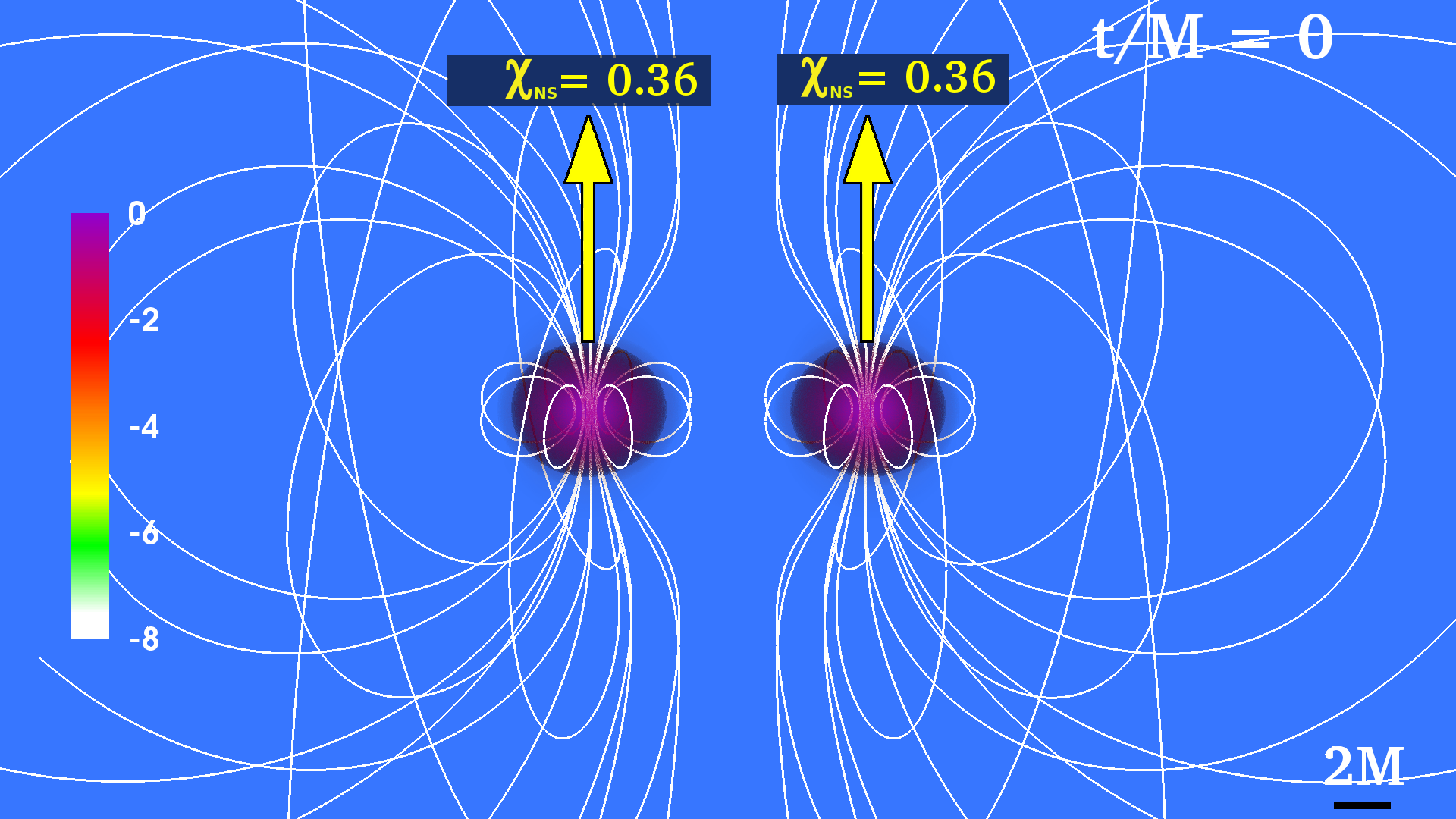}
  \includegraphics[width=0.33\textwidth]{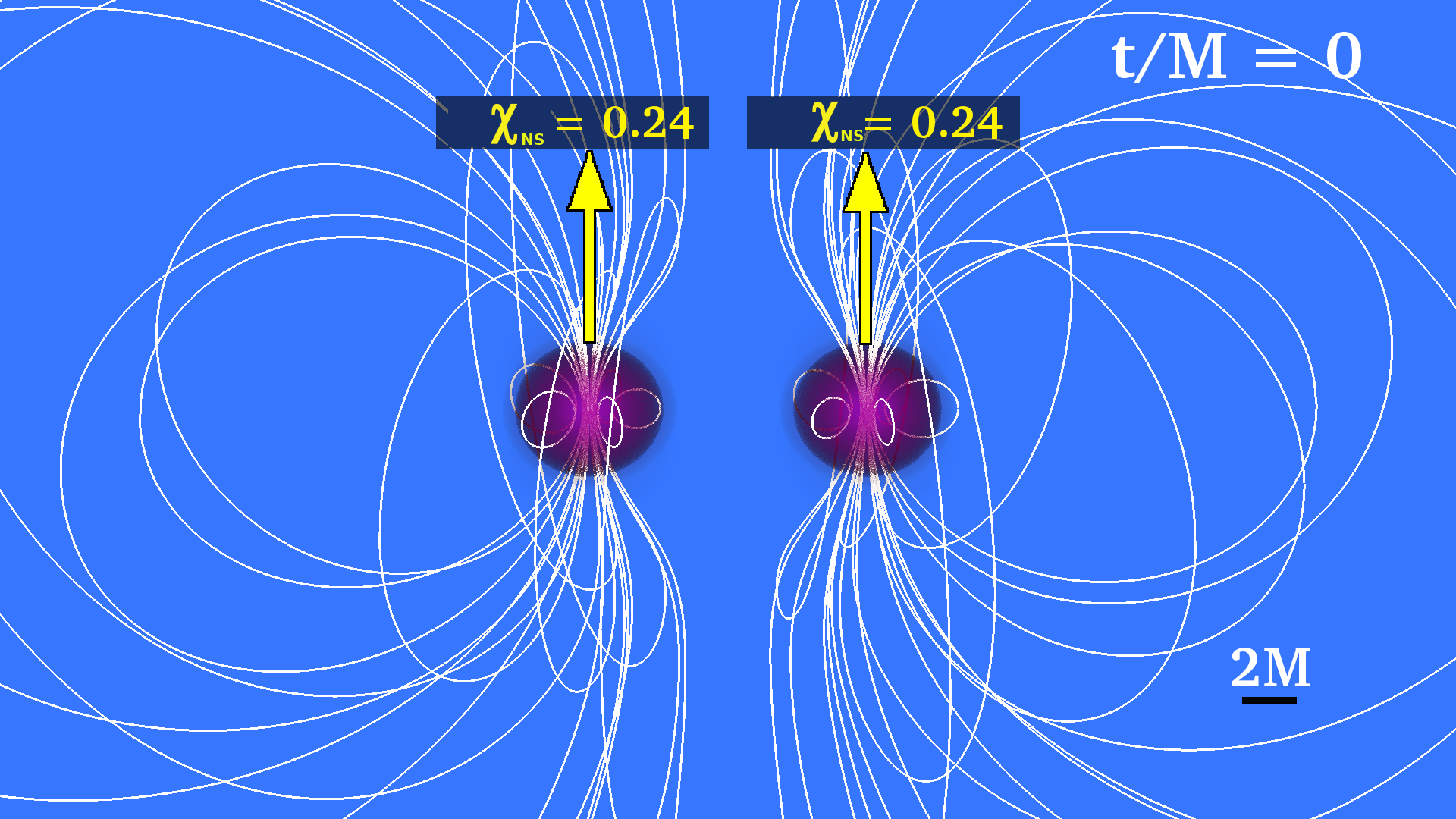}
  \includegraphics[width=0.33\textwidth]{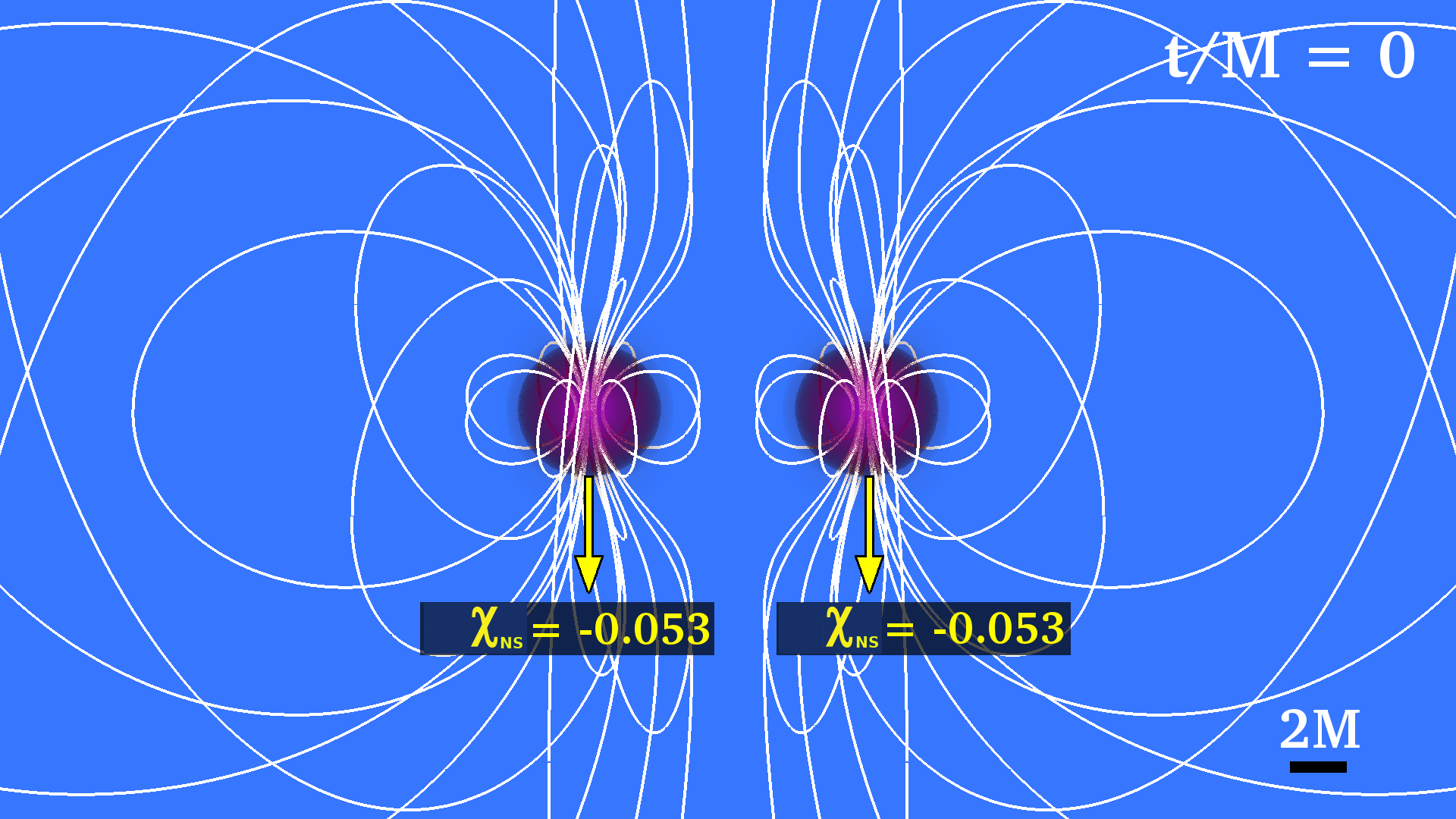}
  \includegraphics[width=0.33\textwidth]{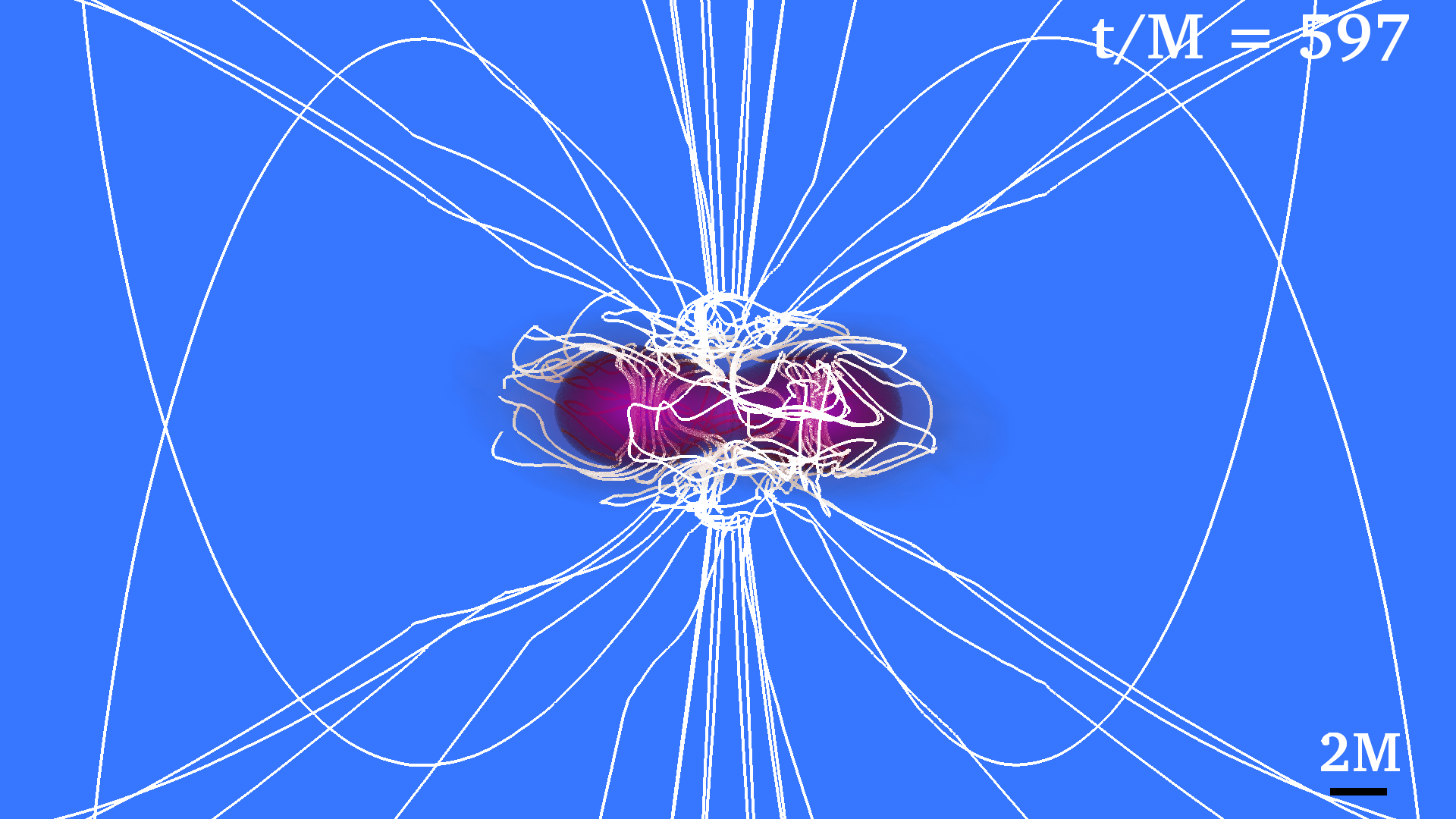}
  \includegraphics[width=0.33\textwidth]{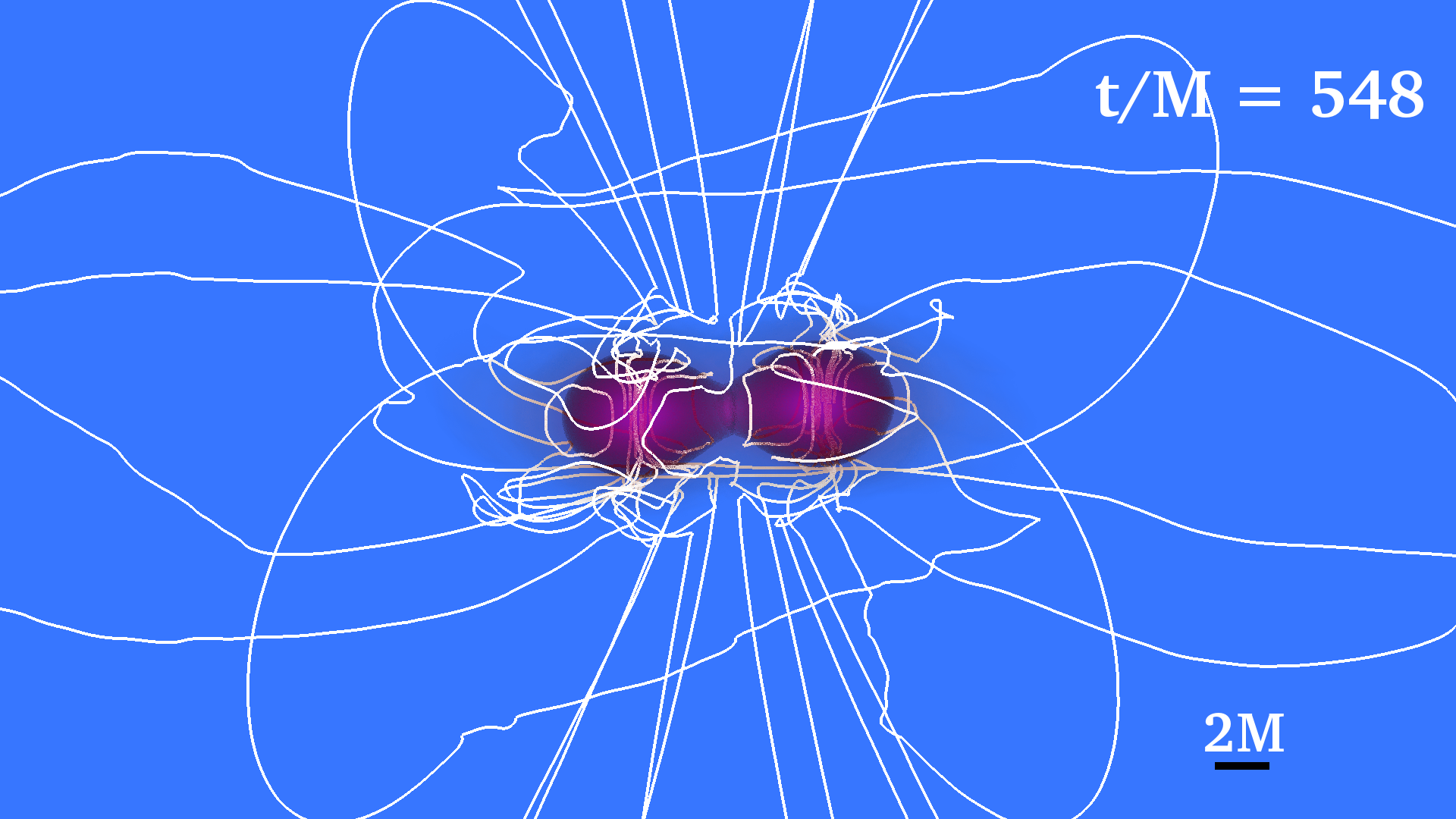}
  \includegraphics[width=0.33\textwidth]{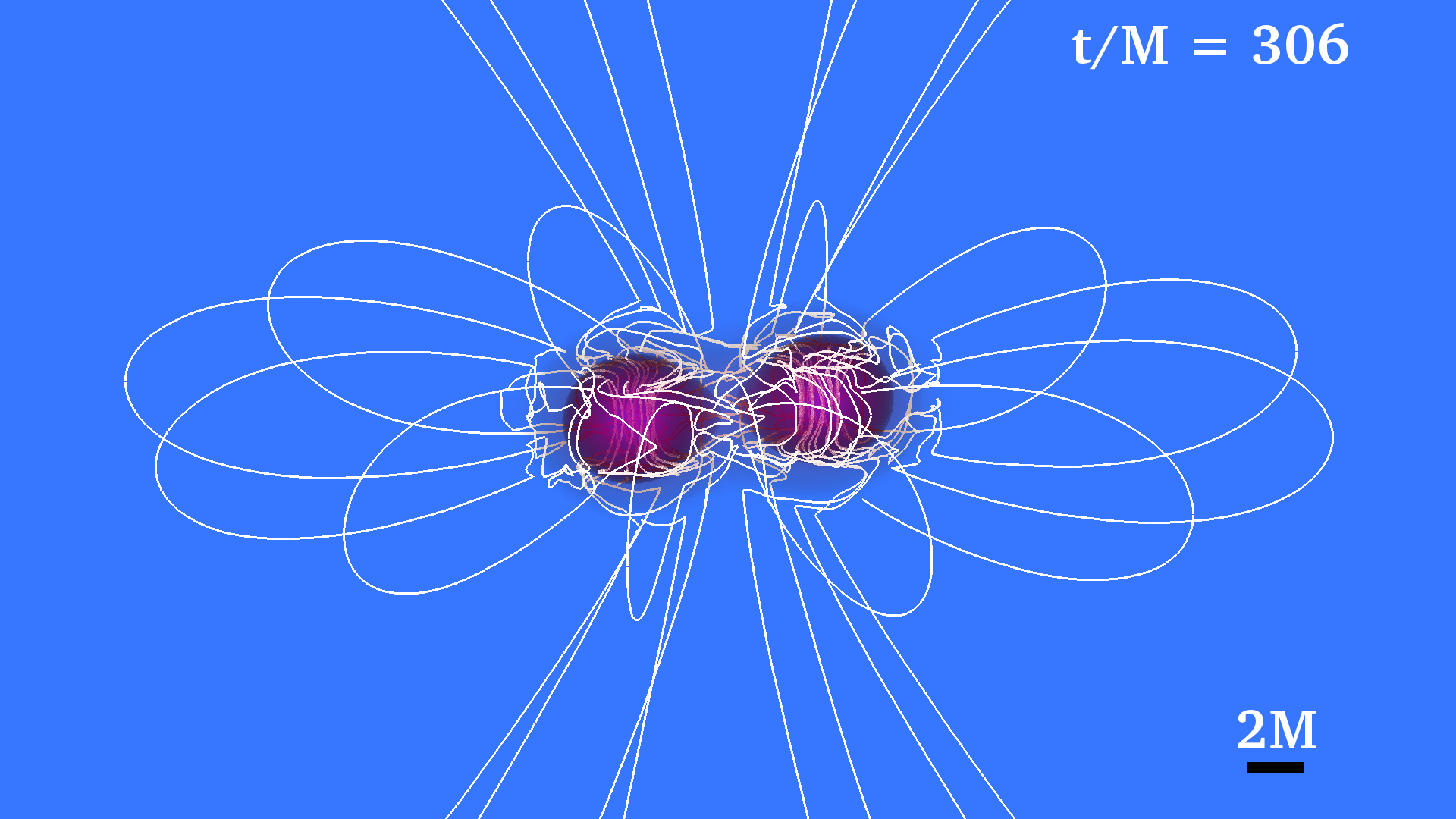}
  \includegraphics[width=0.33\textwidth]{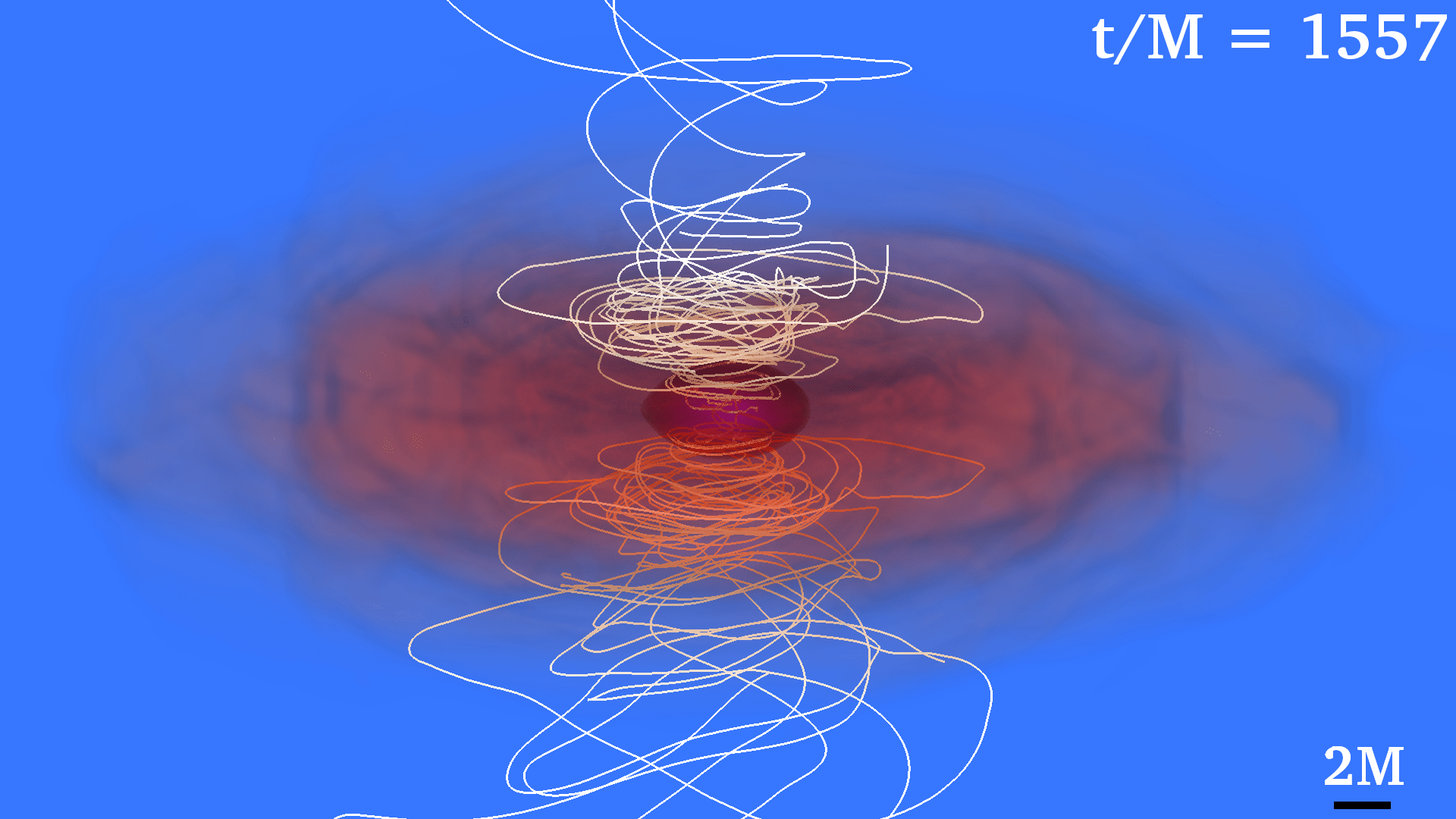}
  \includegraphics[width=0.33\textwidth]{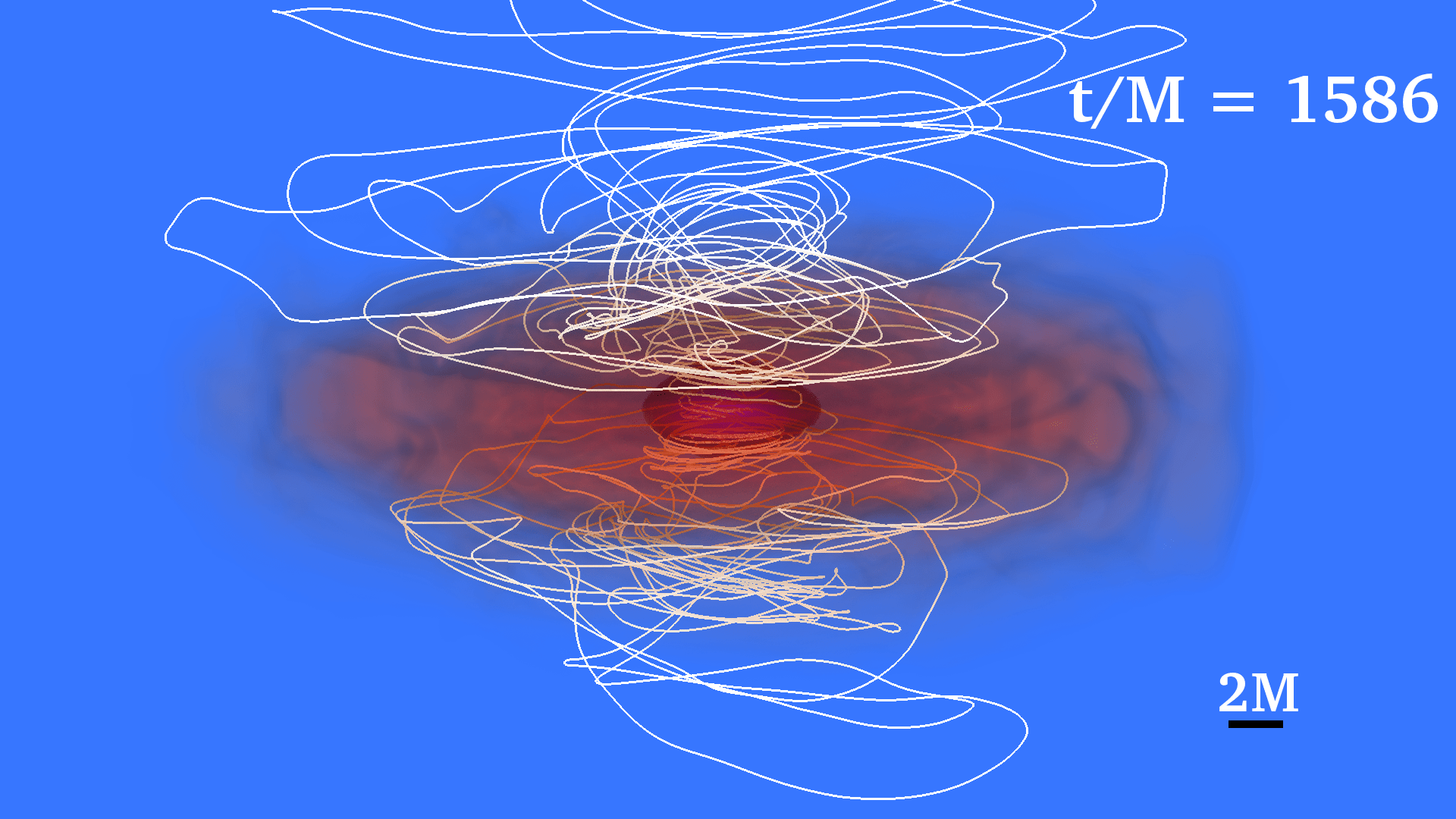}
  \includegraphics[width=0.33\textwidth]{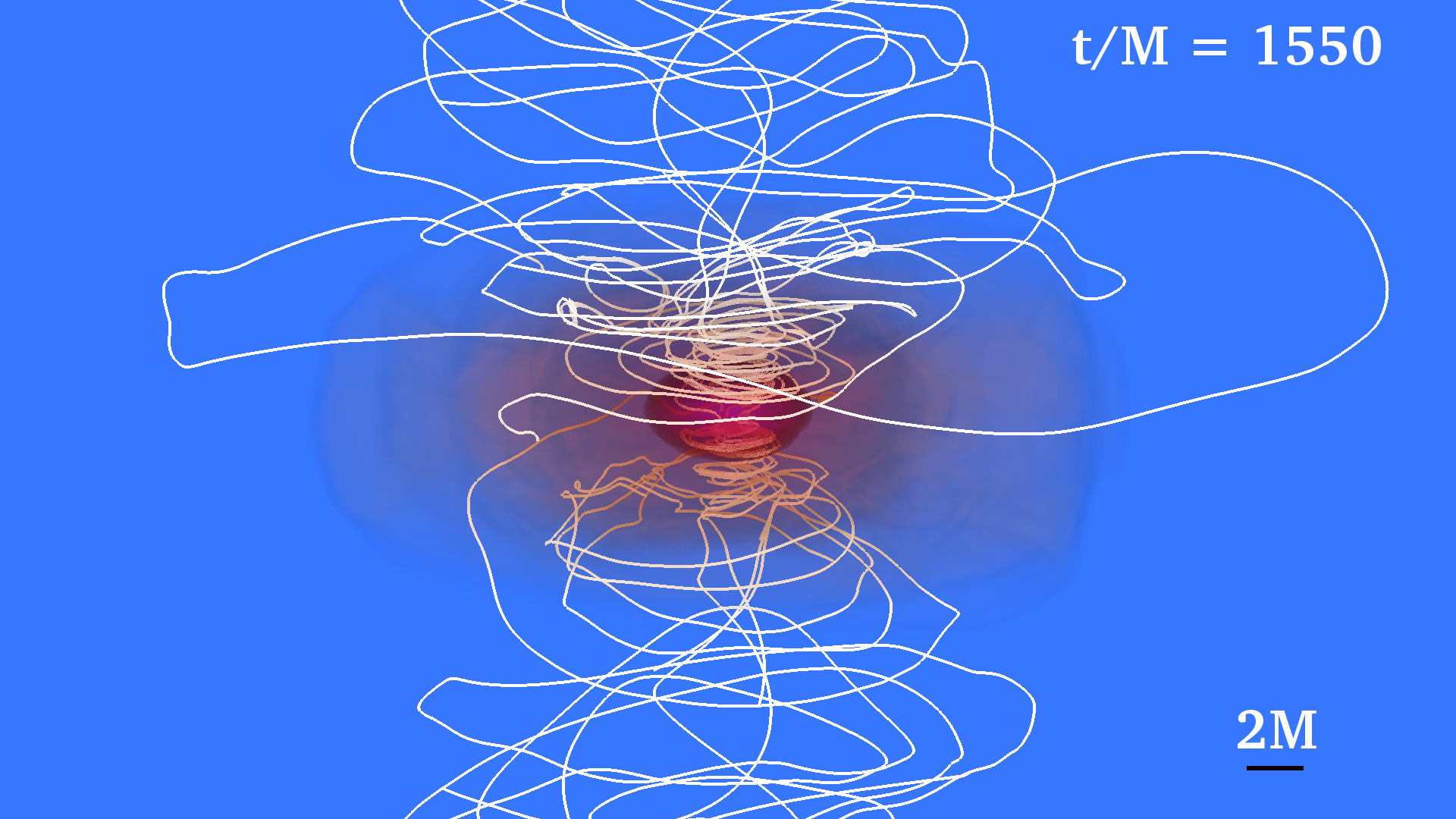}
  \includegraphics[width=0.33\textwidth]{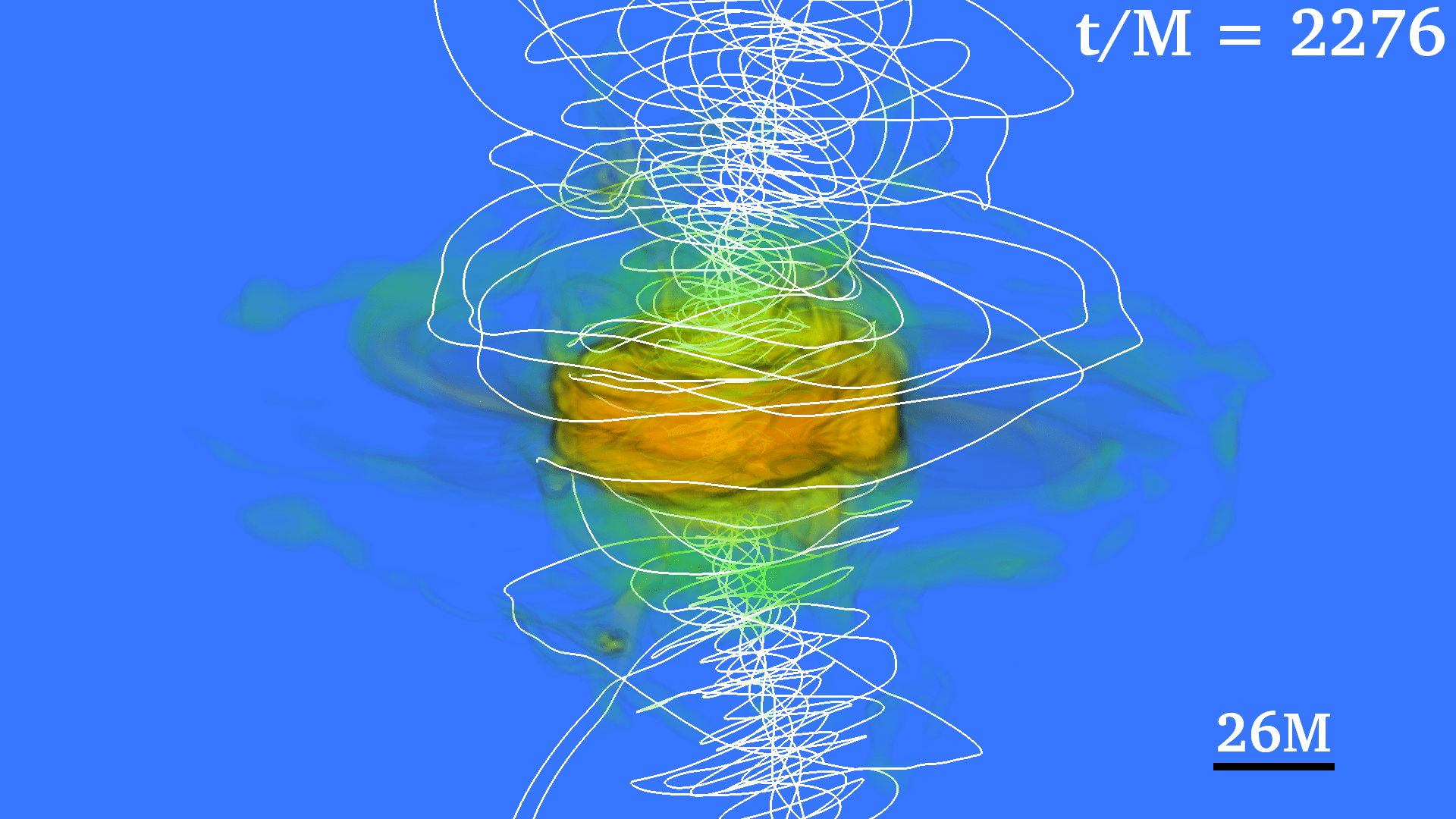}
  \includegraphics[width=0.33\textwidth]{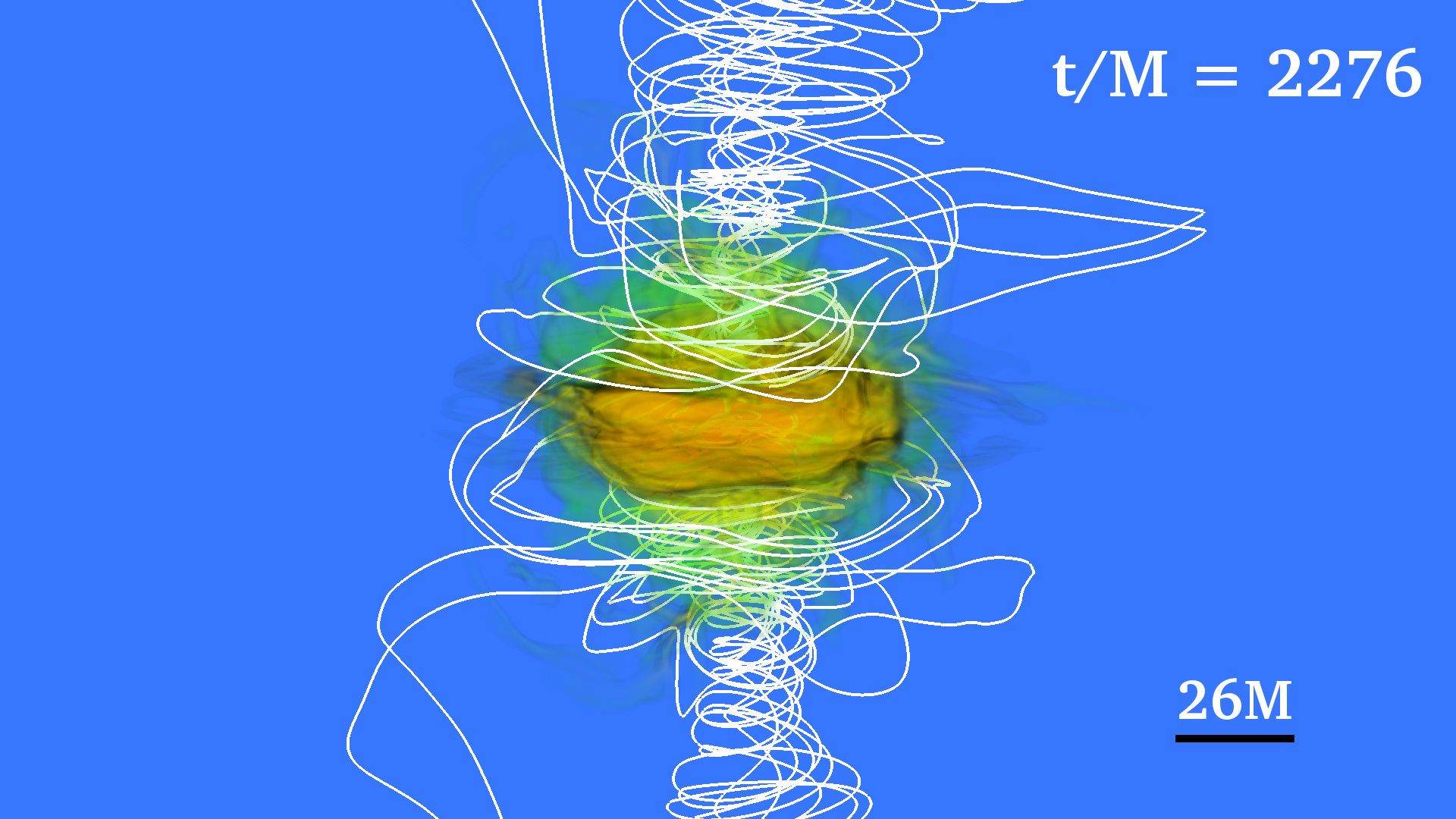}
  \includegraphics[width=0.33\textwidth]{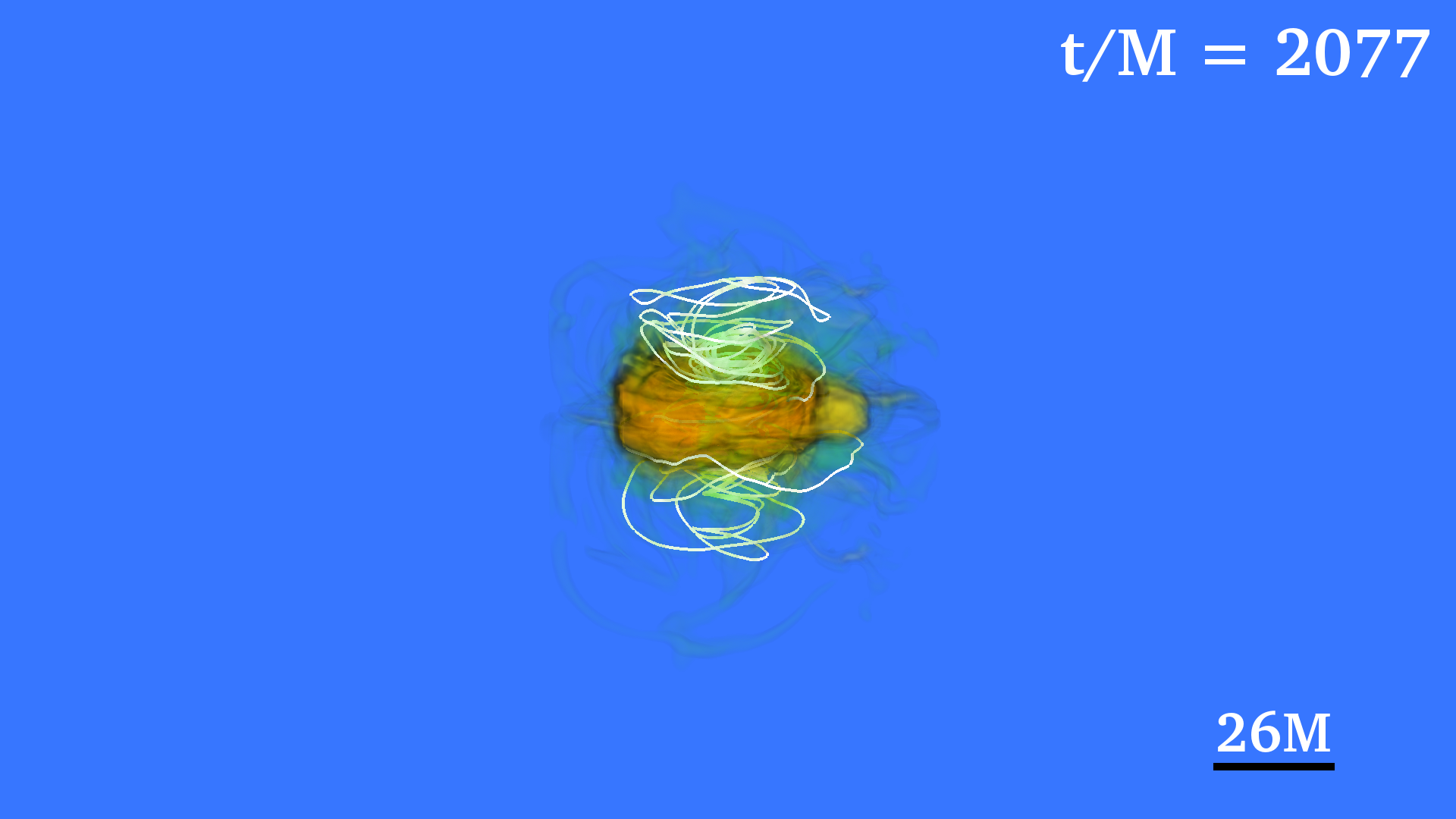}
  \includegraphics[width=0.33\textwidth]{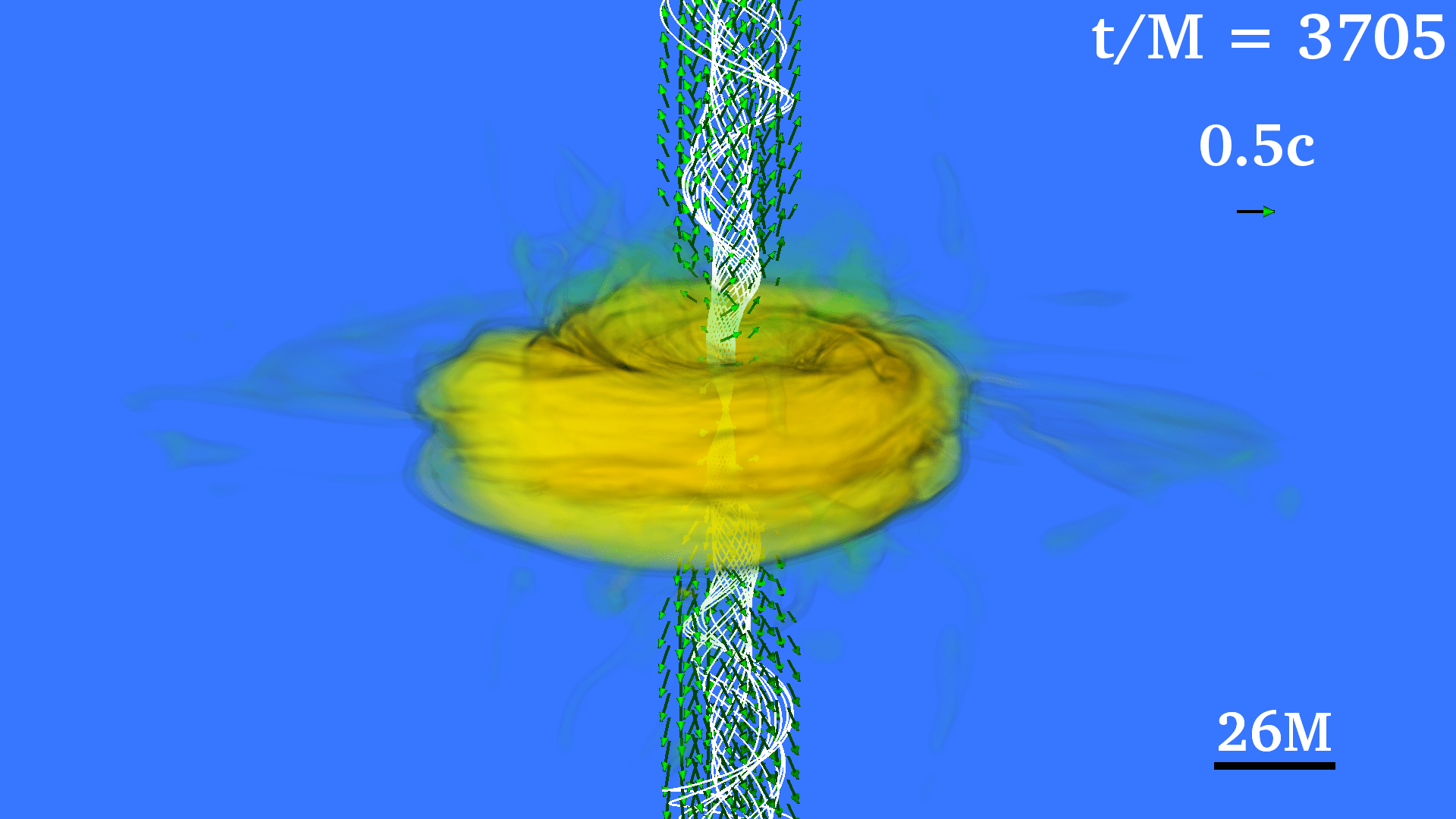}
  \includegraphics[width=0.33\textwidth]{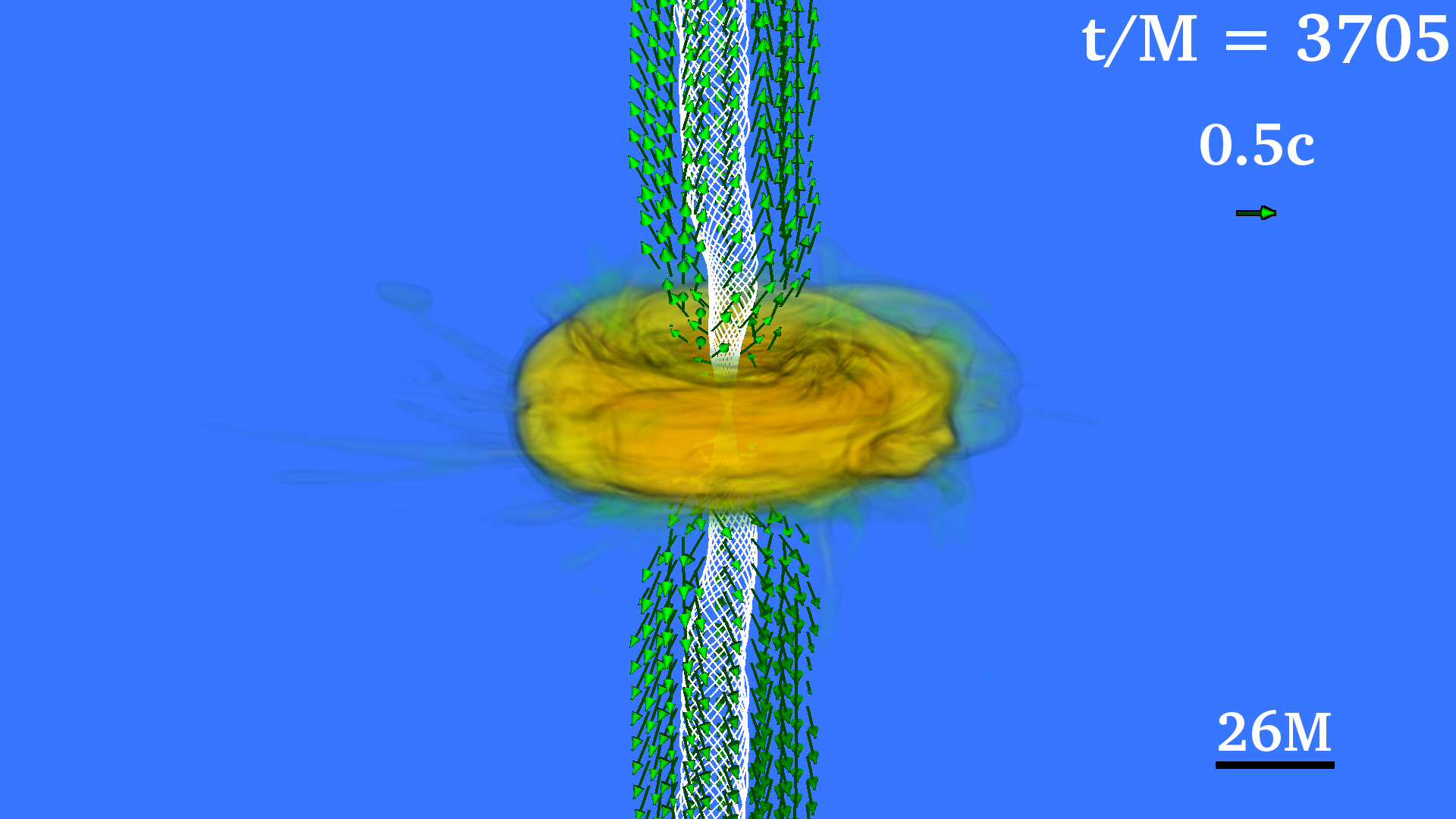}
  \includegraphics[width=0.33\textwidth]{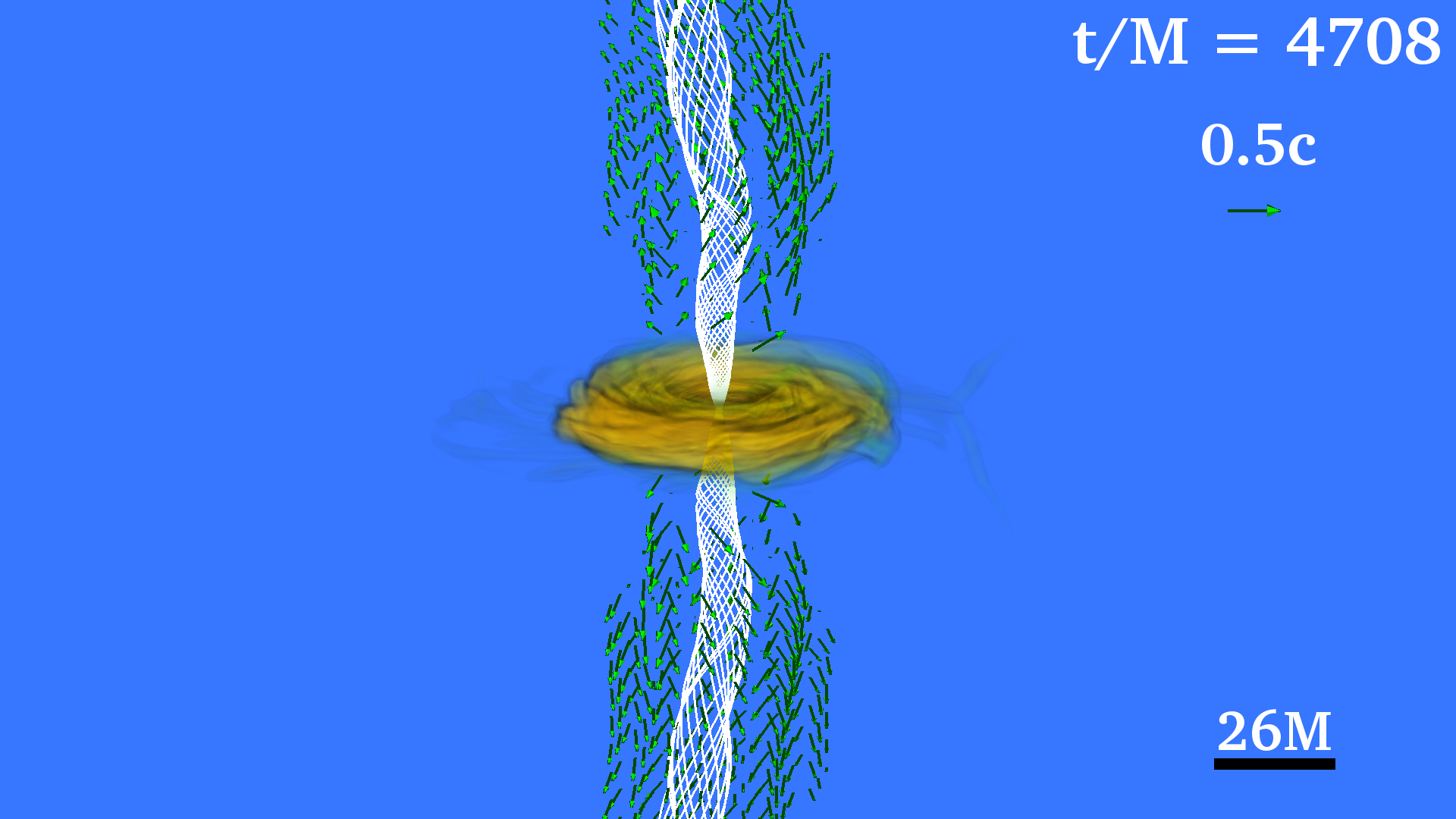}
  \includegraphics[width=0.33\textwidth]{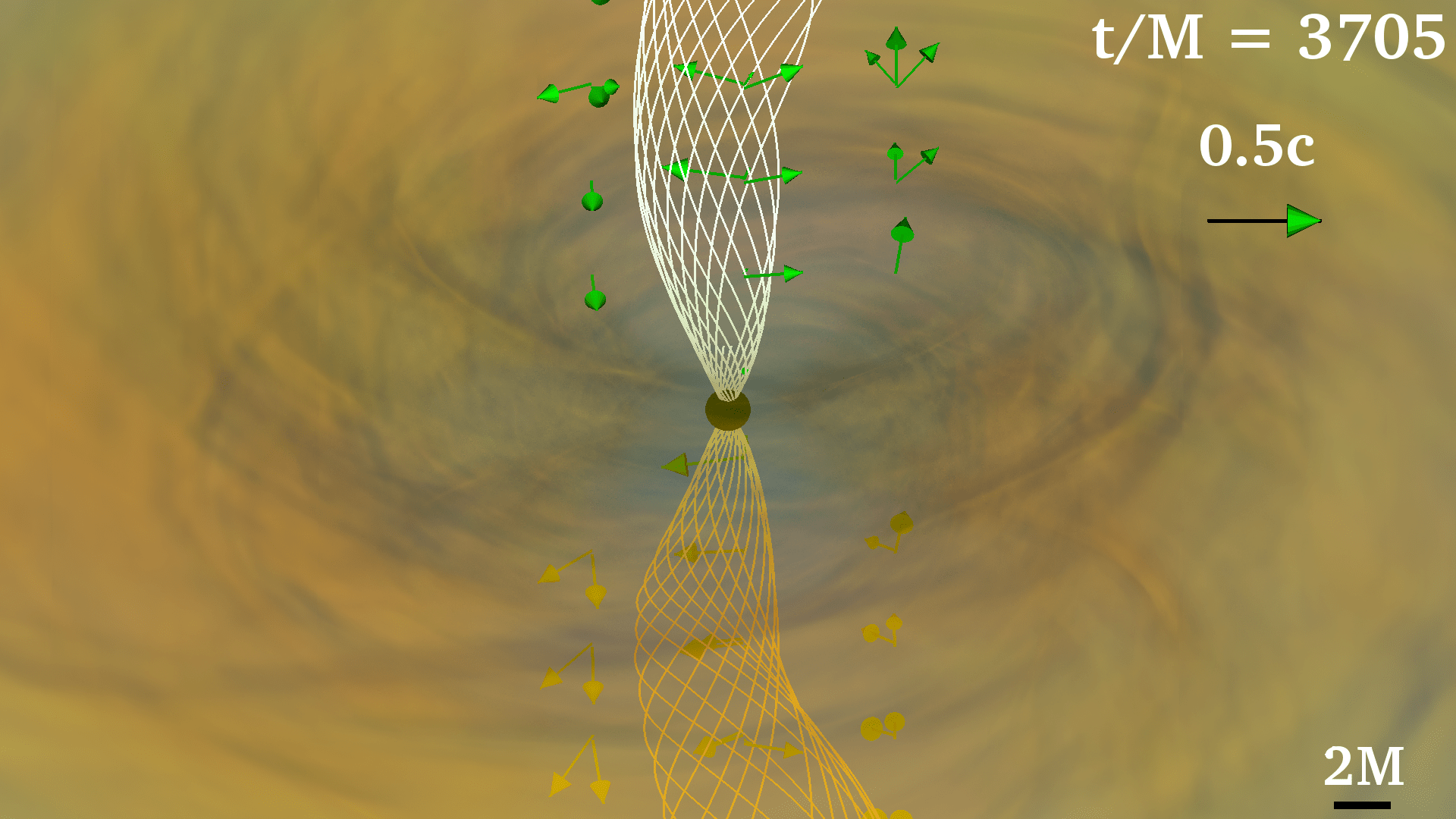}
  \includegraphics[width=0.33\textwidth]{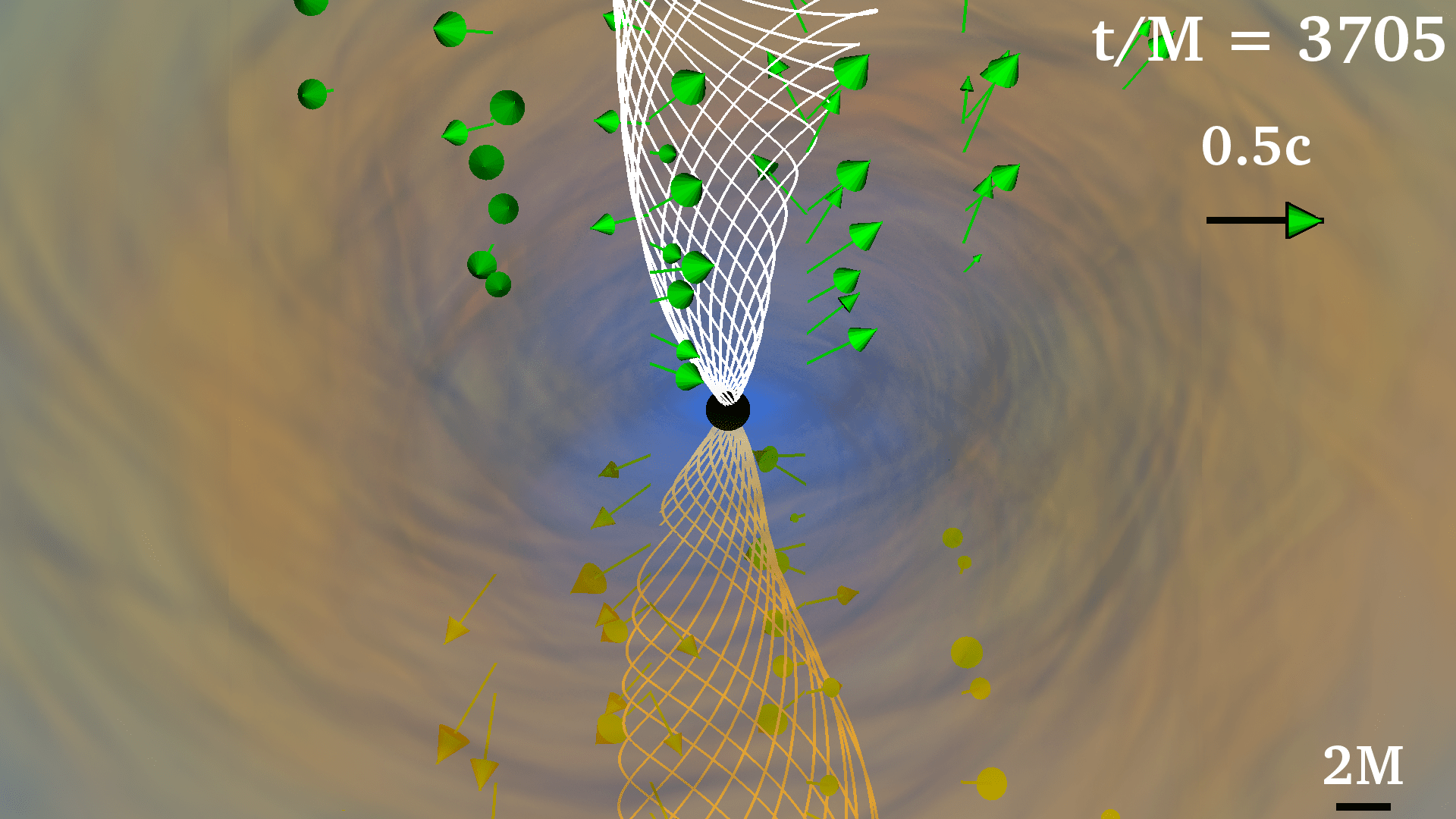}
  \includegraphics[width=0.33\textwidth]{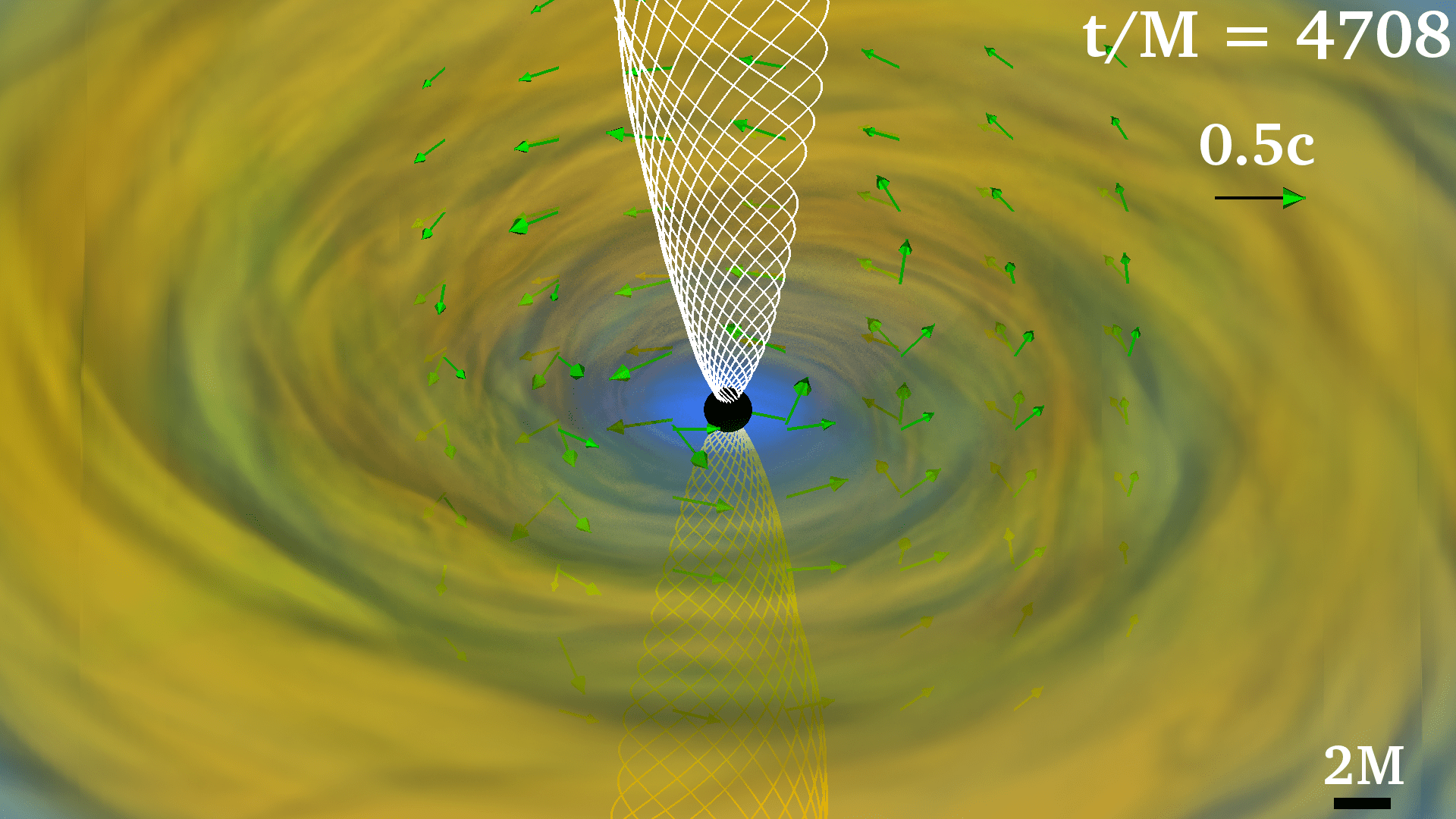}
  \caption{Volume rendering of rest-mass density $\rho_0$, normalized to the initial maximum
    value $\rho_{0,\text{max}}\simeq 10^{14.4}(1.625\,M_\odot/M_{\rm NS})^2\text{g/cm}^3$
    (log scale), at selected times for magnetized cases (see Table~\ref{table:summary_NSNSresults}):
    Msp0.36 (left column), Msp0.24 (middle column), and Msm0.05 (right column). See also Fig.~1~in
    \cite{Ruiz:2016rai}. Bottom panels highlight
    the system after an incipient jet is launched. Arrows indicate plasma velocities while white lines
    show the magnetic field structure. The BH apparent horizon is displayed as a black sphere (see
    bottom panels). Here $M=1.47\times 10^{-2}(M_{\rm NS}/1.625M_\odot)\rm ms$ = $4.43(M_{\rm
      NS}/1.625M_\odot)\rm km$.
    \label{fig:NSNS_snap}}
\end{figure*}

Following~\cite{Ruiz:2016rai}, to reliably evolve  the  exterior magnetic field  and, at the same
time, mimic magnetic-pressure dominance that characterizes the likely force-free, pulsar-like exterior
magnetosphere, we initially enforced a low and variable density in regions where
magnetic field stresses dominate over the fluid pressure gradient (see~Eq.~4~in~\cite{Ruiz:2018wah}),
such that the magnetic-to-gas-pressure ratio in the NS exterior is $\beta^{-1}_{\rm ext}=100$, which
increases the total rest-mass of the system by $\lesssim 0.5\%$. 
For the subsequent evolution, we integrate  the ideal GRMHD equations everywhere imposing a
density floor in regions where $\rho_0^{\rm atm}\leq 10^{-10} \rho_0^{max}$, where
$\rho_0^{max}$ is  the initial maximum density of the NS, as is typically done in hydrodynamics
schemes to recover the so-called primitive variables~(see e.g.~\cite{Font:2007zz}).
%
\subsection{Grid structure}
\label{subsec:grid}
The grid hierarchy used in our simulations is summarized in Table~\ref{table:grid}. It consists
of  two sets of seven nested refinement boxes, the innermost ones centered on each star. Once they overlap
they are replaced by a common box centered on the system center of mass. Each set consists of
seven boxes that differ in size and in resolution by factors of two. The finest box around the star
has a~side half-length of $\sim 1.3\,R_{\rm NS}$, where $R_{\rm NS}$ is the initial NS equatorial
radius~(see Table~\ref{table:NSNS_ID}). In all cases, the initial NS radius is resolved  by $\sim 66$
grid points. We impose reflection symmetry across the orbital plane. Note that this resolution
matches the medium resolution used in~\cite{Ruiz:2016rai}, although in terms of grid points per
NS radius the resolution used here is slightly larger.
%
\subsection{Diagnostics}
\label{subsec:diagnostics}
To verify how close to equilibrium our initial NSNS configurations are, we monitor the
maximum value of the rest-mass density $\rho_0$ during the early inspiral
and find oscillations with an amplitude
of about $\lesssim 1\%$ even in our highest spinning case ($\chi_{\rm NS}=0.36$), as displayed in
Fig~\ref{fig:rho_max_oscillations}.

On the other hand, to validate our numerical results, we monitor the $L_2$ normalized constraints
computed via Eqs.~(40)-(41)~in~\cite{Etienne:2007jg}. We find that during the inspiral and after
the formation of the HMNS the constraints oscillate around $1\%-2\%$. During NSNS merger and BH
formation, they peak at~$\lesssim 6\%$ in the pure hydrodynamic cases, and at  $\lesssim 8\%$ in
the magnetized cases. After that point, the constraints settle back to~$\lesssim 0.1\%$.

The BH apparent horizon is located and monitored through the {\tt
  AHFinderDirect} thorn \cite{ahfinderdirect}. We estimate the BH mass
$M_{\text{BH}}$ and the BH dimensionless spin parameter $a/M_{\rm BH}$
using the isolated horizon formalism~\cite{dkss03}. To measure the
flux of energy and angular momentum carried away by GWs, we use a
modified version of the {\tt Psikadelia} thorn that computes the Weyl
scalar $\Psi_4$, which is decomposed into $s=-2$ spin-weighted
spherical harmonics~\cite{Ruiz:2007yx} at different radii between
$r_{\rm min}\approx 30M\sim 133(M_{\rm NS}/ 1.625M_\odot)\rm km$ and
$r_{\rm max}\approx 170M\sim 752(M_{\rm NS}/1.625M_\odot)\rm km$.  We
find that between $\sim 0.8\%$ and $\sim 1.4\%$ of the total energy of
our NSNS models is radiated away during the evolution in form of
gravitational radiation, while between $\sim 12\%$ and $\sim 19\%$ of
the angular momentum is radiated (see~Table~\ref{table:summary_NSNSresults}).
The escaping mass (ejecta) is computed via $M_{\rm esc} = \int \rho_*\,d^3x$
outside a coordinate radius $r > r_0$, and under the conditions that: a) $-1 - u_t > 0$
(fluid particle energy per unit rest-mass), and b) the radial velocity of the
ejected material $v^r > 0$. Here  $\rho_*\equiv
-\sqrt{\gamma}\rho_0\,n_\mu\,u^\mu$. Varying $r_0$ between $30M\approx 133
(M_{\rm NS}/1.625M_\odot)\rm km$ and $r_{\rm max}\approx 100M\sim 443 (M_{\rm NS}/
1.625M_\odot)\rm km$, we checked that the ejecta masses we report are independent
  of $r_0$. Between
$0.02\%$ and $0.14\%$ of the total rest-mass of the system is
ejected. Notice that ejecta masses of $\sim 0.1\%$ or greater are required
for detectable kilonovae by current or planned telescopes, such
as~LSST~\cite{Metzger:2011bv}.

We also monitor the conservation of both the interior mass $M_{\rm int}$ and the interior angular
momentum $J_{\rm int}$  of the system contained in the numerical domain via Eqs. (19)-(22)
in~\cite{Etienne:2011ea}.  These quantities coincide with the ADM mass and ADM angular momentum
of the system at spatial infinity. Taking  into account the GW radiation losses and the escaping
mass we find that, in all the NSNS configurations considered here, the violation of the conservation
of~$M_{\rm int}$~is $\lesssim 1\%$ along the whole evolution, while the violation of the conservation
of~$J_{\rm int}$ is $\sim 1\%$ in the antialigned and irrotational cases, and $\lesssim 4\%$ in
the other cases.
In addition, we monitor the conservation of the rest mass $M_0=\int \rho_* d^3x$, as well as the
magnetic energy growth outside the BH apparent horizon~$\mathcal{M} =\int u^\mu u^\nu T^{(EM)}_{\mu\nu}\,dV$
as measured by a comoving observer~\cite{Ruiz:2017inq}.

To probe MHD turbulence  in our systems, we compute the effective
Shakura--Sunyaev $\alpha_{\rm SS}$ parameter~\cite{Shakura73} associated  with the effective
viscosity due to magnetic stresses through $\alpha_{\rm SS}\sim T^{\rm EM}_{\hat{r}\hat{\phi}}/P$
(see Eq. 26 in~\cite{FASTEST_GROWING_MRI_WAVELENGTH}).  We also  verify that the MRI is
captured in the post-merger phase of our simulations by computing the  quality factor
$Q_{\rm MRI}\equiv\lambda_{\rm MRI}/dx$, which  measures the number of grid points per fastest
growing MRI mode. Here $\lambda_{\rm MRI}$ is  the fastest-growing MRI wavelength defined
as $\lambda_{\rm\tiny{MRI}} \approx 2\,\pi\,{\sqrt{|b_{\tiny{P}}b^{\tiny{P}}|/(b^2+\rho_0\,h)}}/
{|\Omega(r,\theta)|}$ where $\tiny{|b^{\tiny P}| \equiv\tiny{\sqrt{b^2-|b_\mu\,(e_{\tiny \hat\phi}
    )^\mu|^2}}}$, and $(e_{\hat\phi})^\mu$ is the orthonormal vector carried by an observer comoving
with the fluid, $\Omega(r,\theta)$ is the angular velocity of the disk remnant,
and $dx$ is the local grid spacing~\cite{UIUC_PAPER2}. Note that typically capturing MRI requires
$Q_{\rm MRI}\gtrsim 10$ (see e.g.~\cite{Sano:2003bf,Shiokawa:2011ih}). We  also compute the outgoing
EM Poynting luminosity $  L=-\int T^{r(EM)}_t\,\sqrt{-g}\,d\mathcal{S}$ across spherical surfaces of
coordinate radii between $r_{\rm ext}=46 M\simeq 204(M_{\rm NS}/1.625M_\odot)\rm km$ and $190 M
\simeq 842(M_{\rm NS}/1.625M_\odot)\rm km$. Finally, we monitor the time and azimuthally averaged
angular velocity $\Omega(t,r)$ of the HMNS in the equatorial plane  as~\cite{Hanauske:2016gia,Kastaun2014}
\begin{equation}
  \Omega(t,r)=\frac{1}{4\pi\,P_c}\int_{t-P_c}^{t+P_c}\int_0^{2\pi}\frac{u^\phi}{u^t}\,dt'\,d\phi\,,
\label{eq:aver_Omega}
\end{equation}
where $u^t$ and $u^\phi$ are components of the four velocity $u^\mu$, and $P_c$ is the period of the
HMNS at birth, the time at which the two dense cores collide (see below).
%
%
\section{Results}
\label{sec:results}
\begin{figure*}
  \centering
  \includegraphics[width=0.49\textwidth]{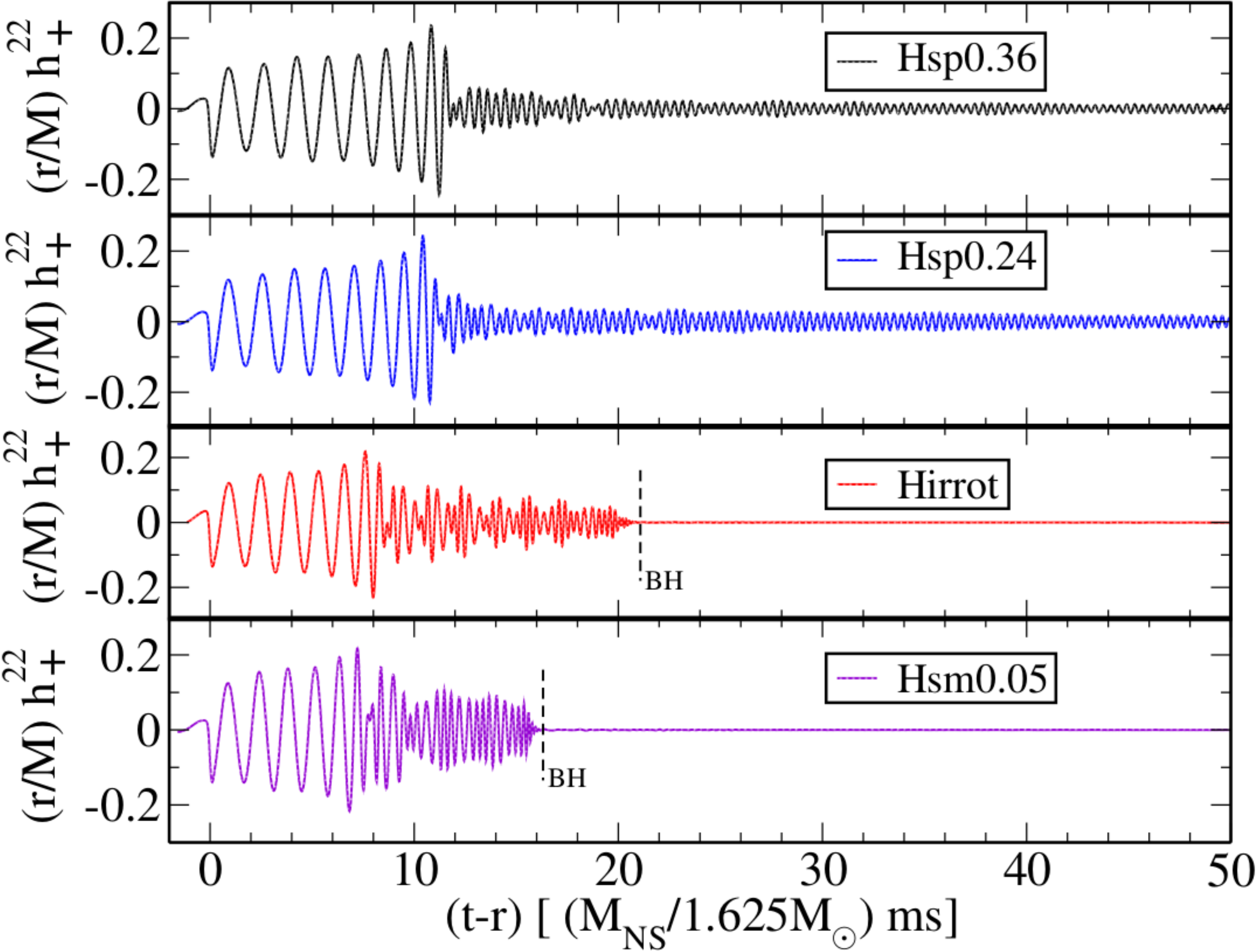}
    \includegraphics[width=0.49\textwidth]{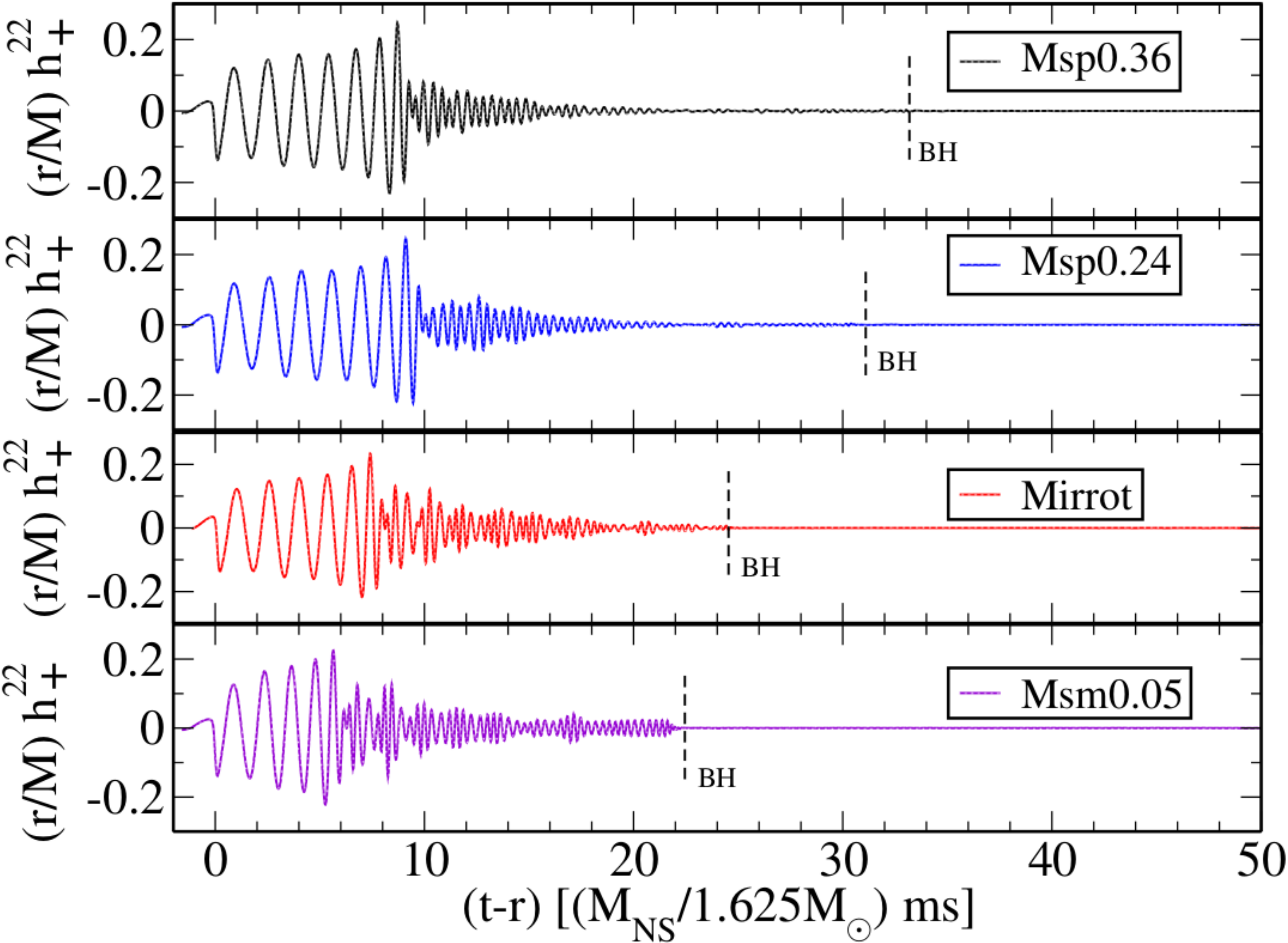}
    \caption{GW strain $h_+^{22}$ (dominant mode) as functions of retarded time, extracted
      at $r_{\rm ext}\approx 100M\sim 443(M_{\rm NS}/1.625M_\odot)\rm km$ for all cases
      listed in Table~\ref{table:summary_NSNSresults}. Left panel displays the GW strain
      in the unmagnetized cases, while right panel displays  the magnetized cases.
      The dashed
      vertical line denotes the BH formation time.
    \label{fig:hydro_magGW}} 
\end{figure*}

The basic evolution and  final outcome of our new spinning NSNS configurations are similar to those
reported in~\cite{Ruiz:2016rai}. The binaries start from an inspiral separation of $45(M_{\rm NS}/
1.625M_\odot)\rm km$ (or $\sim3-4$~orbits before merger, see first column in Fig.
\ref{fig:NSNS_snap_hydro} and first row in~Fig. \ref{fig:NSNS_snap}). As the GWs extract energy and
angular momentum, the orbital separation shrinks and, depending on the initial spin of the NSs, after
about $410M-750M\sim 8(M_{\rm NS}/1.625M_\odot){\rm ms}-11(M_{\rm NS}/1.625M_\odot)\rm ms$ the stars
merge (see second column in see in Fig.~\ref{fig:NSNS_snap_hydro} and second row in Fig.
\ref{fig:NSNS_snap}). We define the merger time $t_{\rm mer}$ as the time of peak amplitude of GWs
(see Fig.~\ref{fig:hydro_magGW}). Following merger, a massive remnant forms with two dense
cores rotating about each other that eventually collide and become a highly differentially
rotating HMNS (see Fig.~\ref{fig:Omega_plot}) wrapped in a dense cloud of matter (see right top panel
in Fig.~\ref{fig:NSNS_snap_hydro} and third row in Fig.~\ref{fig:NSNS_snap}). The HMNS is composed
by matter with a  rest-mass density~$\rho_0\gtrsim 10^{13.5}(1.625\,M_\odot/M_{\rm NS})^2\text{g/cm}^3$.
So,  as it shown in the third row of Fig.~\ref{fig:NSNS_snap}, the larger the initial
spin of the NSs the denser the matter wrapping around the new-born HMNS.
The HMNS has an initial coordinate equatorial radius of roughly $R_{\rm eq}\sim 3.5M\sim 15.5(M_{\rm NS}/1.625M_\odot)
\rm km$, polar radius of $R_{\rm pol}\sim 1.5M\sim 6.7(M_{\rm NS}/1.625M_\odot)\rm km$, and a rest mass of $M_0\simeq
3.2M_\odot(M_{\rm NS}/1.625M_\odot)$ that exceeds the 
supramassive limit, i.e. the maximum value
allowed for uniformly rotating stars with a $\Gamma=2$ polytropic EOS, i.e. $M_0\simeq 2.4(M_{\rm NS}/1.625
M_\odot)M_\odot$~\cite{Cook:1993qj,LBS2003ApJ}. Redistribution of the angular momentum triggered by 
torques induced by the nonaxisymmetric matter distributions in the star~\cite{Kiuchi:2017zzg}, magnetic
winding~\cite{Shapiro:2000zh} and/or magnetically-driven instabilities~\cite{kskstw15,Sun:2018gcl}, along
with dissipation of angular momentum due to gravitational radiation, cause the HMNS to undergo delayed
collapse to a BH (see bottom panel in Fig.~\ref{fig:NSNS_snap_hydro} and forth row in Fig.~\ref{fig:NSNS_snap}).
By contrast to the unmagnetized cases, where the nascent, highly spinning BH + disk
configuration simply settles down (see~bottom panel in~Fig.~\ref{fig:NSNS_snap_hydro}),
in all the magnetized cases the spinning BH + disk remnant is an engine that launches a
magnetically sustained
jet whose outgoing EM Poynting  luminosity is consistent with sGRBs (see bottom panels
in~\ref{fig:NSNS_snap} and Fig. I in~\cite{Ruiz:2016rai}). 

In the following section, we describe the final outcome of our NSNS mergers that differ in magnetic field
content (unmagnetized and pulsar-like magnetized cases), and in the initial spin of the NSs ($\chi_{\rm NS}=
-0.05$, $0.24$, and $0.36$). For completeness, we also include the irrotational P-case already reported
in~\cite{Ruiz:2016rai}. Key results from our models are displayed in Table~\ref{table:summary_NSNSresults}.
%
%
%
\subsection{Unmagnetized NSNS binaries}
\label{sec:hydrocase}
As in spinning binary BH mergers, the magnitude and direction of the
initial spin of the NSs with respect to the orbital angular momentum
affects the merger time $t_{\rm mer}$.  Left panel of
Fig.~\ref{fig:hydro_magGW} displays the GW strain of the dominant mode
$h^{22}_+$ as a function of the retarded time at a coordinate
extraction radius of~$r_{ext}\approx 100 \sim 443 (M_{\rm
  NS}/1.625M_\odot)\rm km$ for all cases.  The NSNS configurations
with spins aligned with the orbital angular momentum (top and second
left panels in Fig.~\ref{fig:hydro_magGW}) undergo about one more
orbit compared to the antialigned and the irrotational cases (third
and bottom panel in Fig.~\ref{fig:hydro_magGW}), which take around
$\sim 3$~orbits to merge (see
Table~\ref{table:summary_NSNSresults}). This so-called orbital hang-up
effect is attributed to the spin-orbit
coupling~\cite{Campanelli:2006uy}.

\begin{figure*}
  \centering
  \includegraphics[width=0.96\textwidth]{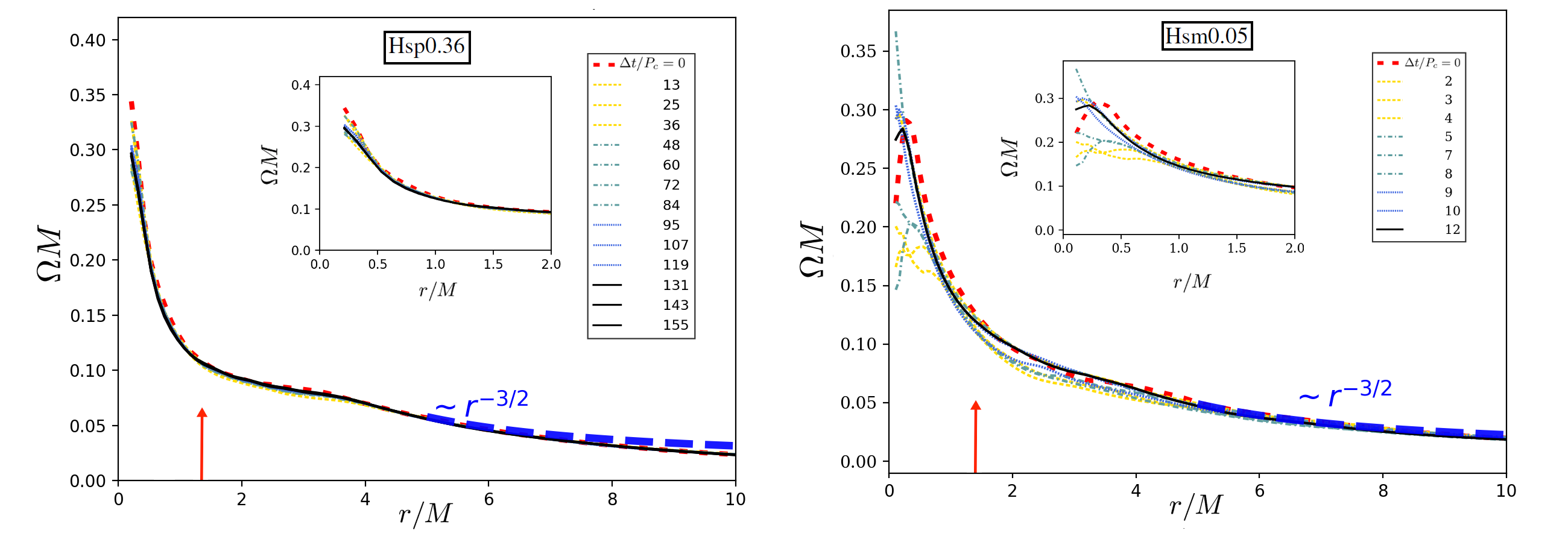}
  \includegraphics[width=0.96\textwidth]{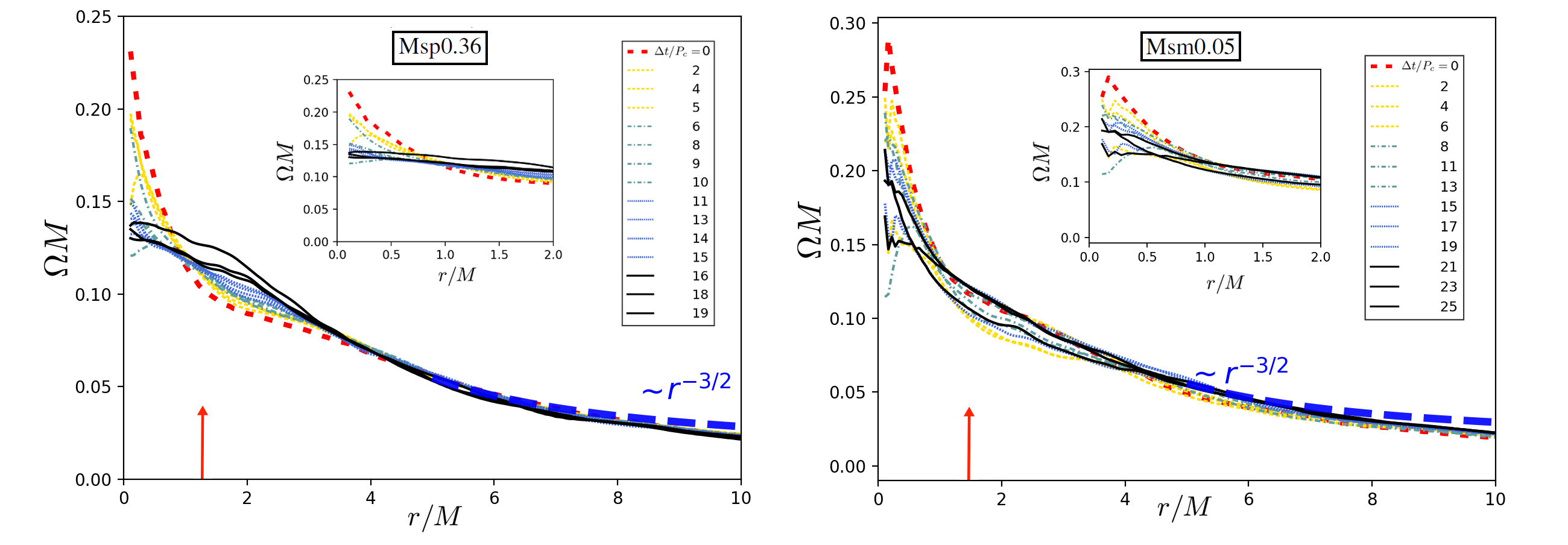}  
  \caption{
    Angular velocity profile of the HMNS in the equatorial plane at $\Delta t = t-t_{\rm HMNS}$,
    with $t_{\rm HMNS}$ the HMNS formation time, and $P_c$ the central HMNS period at $t=t_{\rm HMNS}$
    (see~Eq.~\ref{eq:aver_Omega}) for our extreme
    cases in~Table~\ref{table:summary_NSNSresults} (unmagnetized and magnetized cases are shown in top and
    bottom panels, respectively). The initial differential rotation profile  is displayed by the red dashed curve, while
    the final profile is shown by the continuous black curve. The thick, blue dashed curve shows a Keplerian angular
    velocity. The arrow denotes the coordinate radius that contains $\sim 50\%$ of the rest-mass of the HMNS~(see Fig.
    \ref{fig:NSNS_snap_hydro} and~\ref{fig:NSNS_snap}). The inset displays the angular velocity in the inner layers of
    the HMNS.
    \label{fig:Omega_plot}}
\end{figure*}

Following merger, a highly deformed HMNS with a rest-mass $M_0\simeq
3.2(M_{\rm NS}/1.625 M_\odot)M_\odot$ is formed spinning with a
central rotation period that ranges between $\sim 0.24(M_{\rm
  NS}/1.625M_\odot)\rm ms$, in case Hsp0.36, to $\sim 0.34(M_{\rm
  NS}/1.625M_\odot)\rm ms$, in case Hsm0.05 (see top panel in
Fig~\ref{fig:Omega_plot}). As thermal pressure during the merger is
not sufficient~\cite{ShiU00}, and the rest-mass of the HMNS exceeds
the maximum allowed for an uniformly rotating cold star by $\sim
33\%$, it can survive only as long as differential rotation in the
bulk of the star is maintained.

\begin{figure}
  \centering
  \includegraphics[width=0.50\textwidth]{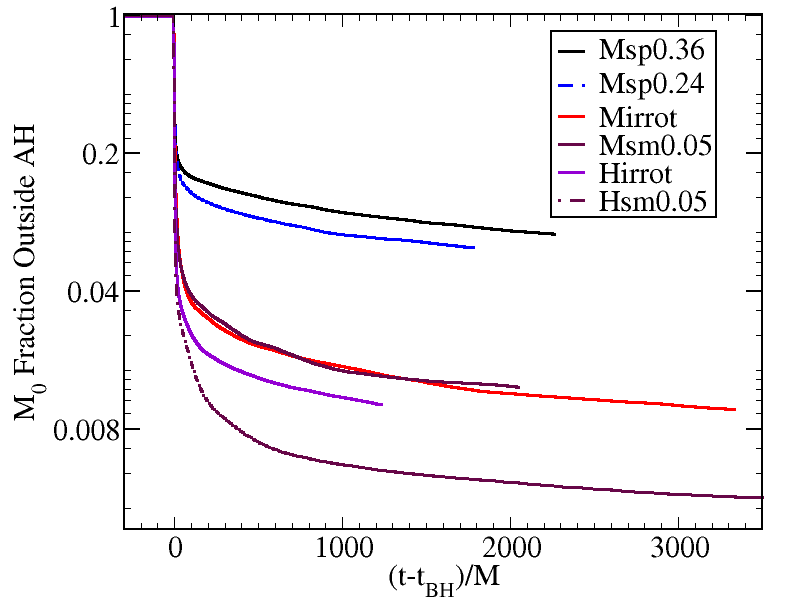}
  \caption{Rest-mass fraction outside the BH apparent horizon  versus time for all cases listed in
    Table~\ref{table:summary_NSNSresults}. The coordinate time has been shifted
    to the BH formation time $t_{\rm BH}$.
    \label{fig:M0_outside}}
\end{figure}

Nonaxisimmetric matter distributions in the HMNS induce the emission of quasiperiodic GWs and
loss of angular momentum~(left panel of Fig.~\ref{fig:hydro_magGW}). Dissipation of angular momentum due to
gravitational radiation is more efficient in cases Hsm0.05 and Hirrot (those with less centrifugal support)
than in cases Hsp0.24 or Hsp0.36. In the first two cases we find that $\gtrsim 17\%$ of the total angular
momentum is radiated away, while in the two aligned cases it is~$\lesssim 13.5\%$ (see~Table
\ref{table:summary_NSNSresults}). Top panel of Fig.~\ref{fig:Omega_plot} shows the averaged angular velocity
profile of the HMNS in cases Hsm0.05 and Hsp0.36 (our extreme cases, see~Table~\ref{table:NSNS_ID}) at different
times and within time intervals of length $\Delta t= \pm\,P_{c}$ about $t$, where $P_c$ is its period at birth
(see Eq.~\ref{eq:aver_Omega}).  In case Hsm0.05 (similar behavior is observed in case Hirrot), angular
momentum is transported from the inner layers of the HMNS to the outer regions (see top right panel in Fig.
\ref{fig:Omega_plot}) triggered by torques arising from the nonaxisymmetric structure in the new-born HMNS
\cite{Kiuchi:2017zzg}. We observe that by $t-t_{\rm mer}\simeq 920M\sim 13.5(M_{\rm NS}/1.625M_\odot)\rm ms$
the HMNS remnant collapses to a BH (see bottom right panel in~Fig.~\ref{fig:NSNS_snap_hydro})
with a mass of $\sim 2.85(M_{\rm NS}/1.625M_\odot)M_\odot$ and spin $a/M_{\rm BH}\sim 0.76$ 
surrounded by a tiny accretion disk (see~Fig.~\ref{fig:M0_outside}). By $t-t_{\rm BH}\sim 900M\sim 13(M_{\rm NS}/
1.625M_\odot)\rm ms$  the accretion rate $\dot{M}$ begins to settle into a quasistationary state and slowly
decays thereafter (see~Fig.~\ref{fig:M0_dot}). At $t-t_{\rm BH}\sim 1200M\sim 18(M_{\rm NS}/1.625M_\odot)\rm
ms$ the  accretion rate  is $\dot{M}\sim 0.4M_\odot/s$, and the rest-mass of the disk is $0.02M_\odot(M_{\rm NS}/
1.625M_\odot)$. We estimate then that the disk will be  accreted in $\tau_{\rm disk}\sim M_{\rm disk}/\dot{M}\sim
41\rm ms$ (see Table~\ref{table:summary_NSNSresults} for case Hirrot).

In contrast, in case Hsp0.36  (similar behavior is observed in case Hsp0.24), the high-angular-momentum
matter in the bulk  of the star rapidly drives the HMNS to a quasiaxisymmetric configuration, reducing the
torques that  induce angular momentum transport and GW radiation. The top left panel in Fig.~\ref{fig:Omega_plot}
shows that after the merger, there are no significant changes in the angular velocity.
The new-born HMNS then  quickly settles into a quasistationary configuration and remains in quasiequilibrium
until the termination of our simulations (see top right panel~in~Fig.~\ref{fig:NSNS_snap_hydro}). However,
by~$t-t_{\rm mer}\approx 3800M\sim 56(M_{\rm NS}/1.625M_\odot)\rm ms$ we observe that the minimum value of the
lapse slowly begins to decrease. Therefore, the HMNS remnant may be gradually evolving to the point of onset of
collapse to a BH.

In most cases, therefore, angular momentum redistributions by nonaxisymmetric torques and gravitational radiation
loss are inefficient mechanisms to trigger the collapse of the HMNS remnant. As we will discuss in Sec.~\ref{sec:magnetized},
other mechanisms, such as magnetic braking and turbulent magnetic viscosity, may be needed to damp the
differential rotation in the HMNS and thereby trigger the collapse to a BH~\cite{Sun:2018gcl}.
%
%
\begin{table*}[]
  \begin{center}
    \caption{Final outcomes. Here $t_{\rm mer}$, $t_{\rm BH}$,  $t_{\rm evo}$ are the NSNS merger time,
      the BH formation time measured after merger, and the full evolution time, respectively. All of them
      in  units of $(M_{\rm NS}/1.625M_\odot)\rm ms$. The mass and
      the dimensionless spin parameter of the remnant BH  Both of them 
      are $M_{\rm BH}$ and~$\tilde{a}\equiv a/M_{\rm BH}$, respectively, $M_{\rm disk}$ is the rest-mass of the
      accretion disk near to the end of the simulation, $\dot{M}$ is the rest-mass accretion rate
      computed via Eq. (A11)~in~\cite{Farris:2009mt} in units of $(M_\odot/s)$, $\tau_{\rm disk}\sim M_{\rm disk}/\dot{M}$
      is the disk lifetime (lifetime of the jet, if any) in units of $(M_{\rm NS}/1.625M_\odot)\rm
      s$, $M_{\rm esc}$ denotes the escaping mass, $\Delta \bar{E}_{\rm GW}\equiv\Delta E_{\rm GW}/M_{ADM}$
      and $\Delta \bar{J}_{\rm GW}\equiv\Delta J_{\rm GW}/J_{ADM}$ are the fractions of the total energy and total
      angular momentum carried away by GWs, respectively, $\alpha_{\rm SS}$ is the
      Shakura--Sunyaev viscosity parameter,  $B_{\rm rms}$ denotes the rms value of the magnetic field
      above the BH poles in units of $(1.625M_\odot/M_{\rm NS})$G, $L_{\rm EM}$ is the Poynting
      luminosity in $\rm erg/s$ driven by the incipient jet and $\eta_{\rm eff}=L_{\rm EM}/\dot{M}$ is the jet
      efficiency.  These two last quantities are time-averaged over the last $500M\sim 7.4
      (M_{\rm NS}/1.625M_\odot)\rm ms$ before the termination of our simulations. Finally, $\Gamma_L$ is the
      maximum fluid Lorentz factor near to $t_{\rm evo}$. [N/A] denotes  ``not applicable''.
      \label{table:summary_NSNSresults}}
    \begin{tabular}{ccccccccccccccccc}
      \hline\hline
          {Model} & $t_{\rm mer}$  & $t_{\rm BH}$ &$t_{\rm evo}$  &$M_{\rm BH}$ &  $\tilde{a}$   & $M_{\rm disk}/{M_0}^{(a)}$     & $\dot{M}^{(b)}$  &  $\tau_{\rm disk}$ & $M_{\rm esc}^{(c)}/{M_0}$ & $\Delta \bar{E}_{\rm GW}$ & $\Delta\bar{J}_{\rm GW}$  & $\alpha_{\rm SS}$          &$B_{\rm rms}$      & $L_{\rm EM}$ & $\eta_{\rm eff}$ & $\Gamma_L$ \\
          \hline
          Hsp0.36  & 11.2 &[N/A]& 62 &[N/A]     & [N/A]       & $0.83\%$     & [N/A]   & [N/A] & $10^{-4}\%$  &   $0.8\%$ & $12.9\%$   & [N/A]               &  [N/A]        &   [N/A]      &   [N/A] &   [N/A] \\
          Hsp0.24  & 10.7 &[N/A]& 62 &[N/A]     & [N/A]       & $0.74\%$     & [N/A]   & [N/A] & $10^{-4}\%$  &   $0.9\%$ & $13.5\%$   & [N/A]               &  [N/A]        &  [N/A]       &   [N/A] &   [N/A] \\
          Hsm0.05  & 7.4  &13.5 & 62 &$2.95M_{\odot}$  & 0.76 & $0.55\%$     & 0.44    & 40.6  & $10^{-5}\%$  &   $1.4\%$ & $18.6\%$   & [N/A]               &  [N/A]        &  [N/A]       &   [N/A] &   [N/A] \\
          \hline 
          Msp0.36  &8.5 & 20.5& 62 &$2.75M_{\odot}$ & 0.78  & $7.82\%$     & 2.71    & 138.5 & $0.14\%$     & $0.76\%$& $11.55\%$    & 0.01 - 0.09         &  $10^{15.8}$  & $10^{52.1}$  &    $0.26\%$  & 1.26 \\
          Msp0.24  &7.9 & 20.0& 62 &$2.79M_{\odot}$ & 0.77  &  $6.65\%$    & 2.70    & 118.2 & $0.12\%$     & $0.76\%$& $12.08\%$    & 0.02 - 0.07          &  $10^{15.8}$  & $10^{52.3}$  &   $0.32\%$  &  1.27 \\
          Msm0.05  &5.6 & 15.8& 77 &$2.93M_{\odot}$ & 0.77  &  $1.0\%$     & 0.49    & 97.9  & $0.02\%$     & $1.1\%$ & $14.56\%$    & 0.01 - 0.06         &  $10^{15.7}$  & $10^{51.5}$  &    $0.36\%$   & 1.21 \\
          \hline
          \hline
          Hirrot     &7.7 &15.5&52 &$2.85M_{\odot}$ & 0.78  &  $0.81\%$    & 0.48    & 81.0  & $10^{-5}\%$  & $1.3\%$ & $16.8\%$     & [N/A]               &  [N/A]        &  [N/A]       &   [N/A]   &   [N/A] \\
          Mirrot$^{(d)}$& 6.8 &18.0 &74 &$2.85M_{\odot}$ & 0.8  &  $1.0\%$     & 0.48    & 100.0 & $0.03\%$     &$1.0\%$ & $14.52\%$     & 0.04 - 0.08         &  $10^{16.0}$  & $10^{51.3}$  &  $0.24\%$  &  1.25   \\
          \hline\hline
    \end{tabular}
  \end{center}
  \begin{flushleft}
    $^{(a)}$ $M_0$ denotes the initial total rest-mass of the system.\\
    {{$^{(b)}$ $\dot{M}$ is reported once the accretion begins to settle into a
    quasistationary state. So, we quote the accretion rate at $t-t_{\rm BH}\sim$ $18(M_{\rm NS}/1.625M_\odot)\rm ms$ for cases Hsm0.05 and Hirrot, at
    $t-t_{\rm BH}\sim 15(M_{\rm NS}/1.625M_\odot)$ $\rm ms$ for cases Msp0.36 and Msp0.24,
    and at $t-t_{\rm BH}\sim$ $20 (M_{\rm NS}/1.625M_\odot)\rm ms$ for cases Msm0.05 and Mirrot (see Fig.~\ref{fig:M0_dot}).}}\\
    {{$^{(c)}$ $M_{\rm esc}$ reported near to the termination of our simulations (see inset in Fig.~\ref{fig:Poynting_plot}).}}\\
    $^{(d)}$ P-case configuration reported in~\cite{Ruiz:2016rai}.
  \end{flushleft}
\end{table*} 
%

%
%

%
%
%
\subsection{Magnetized binaries}
\label{sec:magnetized}
As the magnetic-to-gas-pressure ratio in the NS interior is initially small ($\beta^{-1}\sim10^{-3}$,
see Sec.~\ref{subsec:idata}), the inspiral dynamics of the magnetized cases proceed basically unperturbed
by the magnetic field,  though we observe that all the magnetized binaries merge slightly earlier
($\lesssim 200M\sim 3(M_{\rm NS}/1.6M_\odot)\rm ms$) than the corresponding unmagnetized cases (see
right panel of~Fig.~\ref{fig:hydro_magGW}). The stars then simply advect the frozen-in magnetic field
lines during the inspiral~(see~second row of Fig.~\ref{fig:NSNS_snap}), and there is no significant
changes in the magnetic energy~$\mathcal{M}$. Once the NSs make contact, the magnetic energy is steeply
enhanced. By $t-t_{\rm mer}\sim 150M\approx 2.2(M_{\rm NS}/1.6M_\odot)\rm ms$ (the time at which the
two dense cores merge) the initial magnetic energy already has been amplified by a factor of $\sim 15$
in the two aligned cases, and by a factor of $\sim 10$ in the Msm0.05 and the Mirrot cases~(see~Fig.
\ref{fig:EM_outside}). Similar behavior was reported in high resolution simulations where
it was found that the KH instability, developed during the NSs contact and persisting  until the HMNS
settles, boost the strength of the magnetic~field along with magnetic winding and the MRI
\cite{Kiuchi:2014hja,Kiuchi:2015sga}.

Following merger, high angular momentum matter originating in the outer layers of the new-born
HMNS begins to settle in a disk around a central core. As it is shown in the third row of
Fig.~\ref{fig:NSNS_snap}, the mass and size of disk depend strongly on the initial 
spin of the NSs. Simultaneously, the inner layers of the star drag the poloidal magnetic
field lines into a toroidal configuration (magnetic winding). As the strength of the toroidal
magnetic field component is amplified, magnetic stresses increase until they are large
enough to redistribute angular momentum and damp the differential rotation
\cite{Shapiro:2000zh,dlsss06a,Sun:2018gcl}. The winding timescale can be estimated
as (see Eq.~2 in~\cite{Sun:2018gcl})
\begin{eqnarray}
  \label{eq:tau_w}
  \tau_{\rm wind}& \sim &\frac{R}{v_{A}}\sim \\
  &&10{\rm ms}\left(\frac{|B|}{10^{15}G}\right)^{-1}\,
  \left(\frac{R}{10^6\rm cm}\right)\,\left(\frac{\rho}{10^{14} \rm g/cm^3}\right)^{1/2}\,,
  \nonumber
\end{eqnarray}
where $R$ is the characteristic radius of the HMNS and $v_{A}=|B|/\sqrt{4\pi\rho}$  the
Alfv\'en speed, with $|B|$ the strength of the magnetic field and $\rho$ the characteristic
density of the star.

We also note that in the HMNS the wavelength~$\lambda_{\rm MRI}$ of the fastest growing MRI is
resolved by $\gtrsim 10$ grid points and it fits within it (see Fig.~\ref{mri_alig}). Thus, it
is likely that the MRI is
operating during the lifetime of the HMNS. We also compute the effective
Shakura--Sunyaev $\alpha_{\rm SS}$ parameter at $t-t_{\rm mer}\approx 1000 M\sim 14.7(M_{\rm NS}
/1.6M_\odot)\rm ms$. We find that in the star, the value of $\alpha_{SS}$ ranges between
$\sim 0.01$ to $\sim 0.09$ (see Table \ref{table:summary_NSNSresults}). Similar values were
reported in high resolution NSNS mergers in~\cite{Kiuchi:2017zzg}. Therefore, it is expected
that magnetic turbulence is operating and it is sustained during the whole lifetime of the HMNS.

Magnetic turbulence can also redistribute angular momentum and damp the differential rotation
in a turbulent viscous timescale of (see Eq.~7 in~\cite{Sun:2018gcl})
\begin{eqnarray}
  \label{eq:tau_mri}
 \tau_{\rm vis} &\sim&{R^{3/2}}\,{M^{-1/2}\,\alpha_{\rm SS}^{-1}}\sim\\
 &&10{\rm ms}\,\left(\frac{\alpha_{\rm ss}}{10^{-2}}\right)^{-1}\,
 \left(\frac{M}{3.2M_\odot}\right)
  \,\left(\frac{\mathcal{C}}{0.3}\right)^{-3/2}\,,\nonumber     
\end{eqnarray}
where $M$ is the characteristic mass of the HMNS and $\mathcal{C}=M/R_{eq}$ its compaction,
with $R_{eq}$ the equatorial radius of the star. Notice that we have estimated $\tau_{\rm vis}$
using an averaged value of $\alpha_{\rm SS}$ during the whole evolution. However, this timescale
can be ``locally'' as long (short) as  $\tau_{\rm vis}\sim 100\rm ms$ ($\tau_{\rm vis}\sim~1\rm ms$),
see~Table~\ref{table:summary_NSNSresults}. On the other hand,  magnetic turbulence can be suppressed
by numerical diffusion~\cite{Kiuchi:2017zzg,Hawley:2011ApJ,Hawley:2013lga} and, therefore, the value
of $\alpha_{\rm SS}$ in our simulations may be underestimated. Higher resolutions than used here may
be required to properly model magnetic turbulence. Nevertheless, it is expected that in higher
resolution studies the timescale $\tau_{\rm vis}$ is shortened (see~\cite{Kiuchi:2017zzg,
  Hawley:2011ApJ,Hawley:2013lga} for a detailed discussion).

%
\begin{figure}
  \centering
  \includegraphics[width=0.49\textwidth]{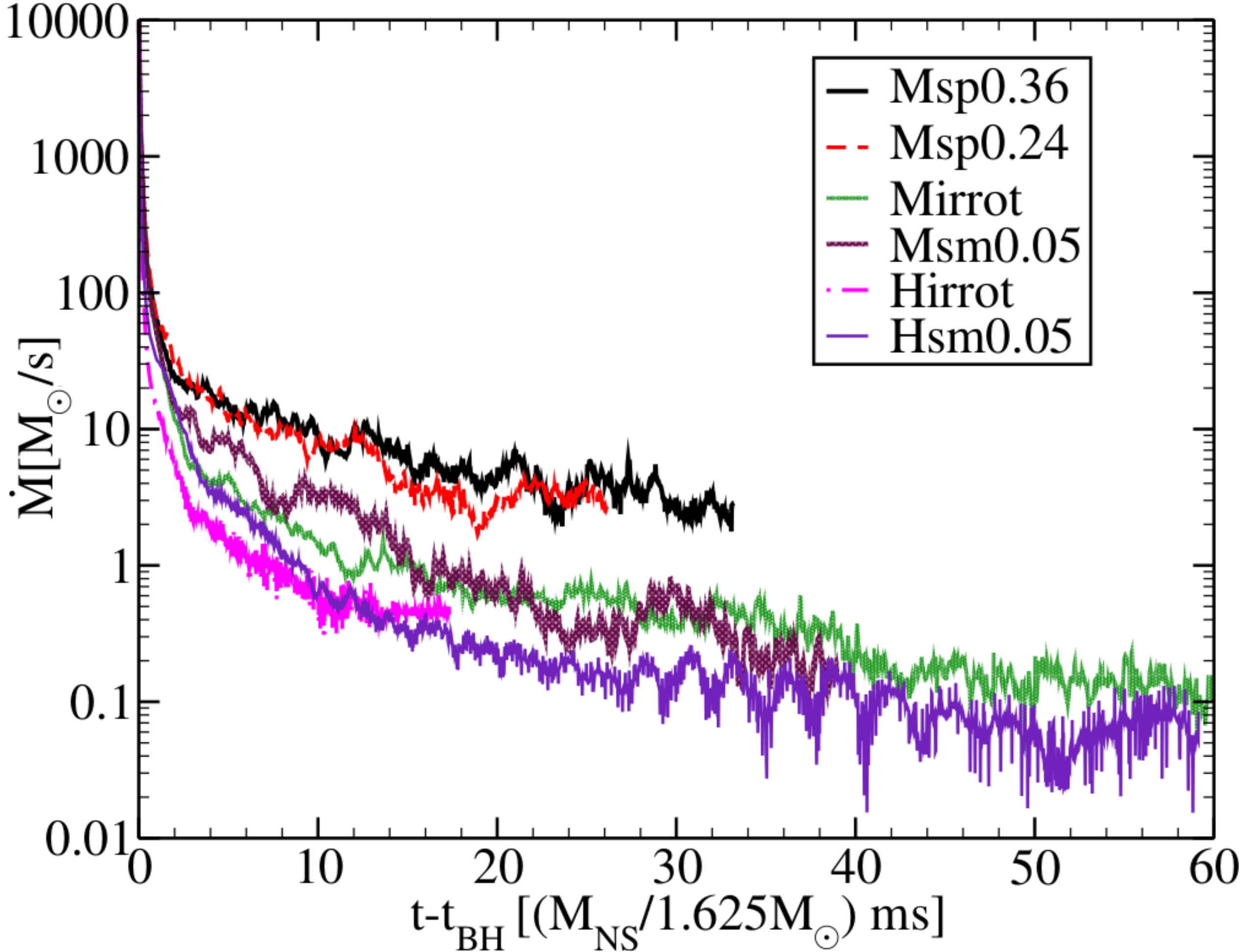}
  \caption{Rest-mass accretion rate for all cases listed in Table~\ref{table:summary_NSNSresults}
    computed via Eq.~(A11) in~\cite{Farris:2009mt}. The coordinate time has been shifted to
    the BH formation time~$t_{\rm BH}$.
    \label{fig:M0_dot}}
\end{figure}
%

\begin{figure}
  \centering
  \includegraphics[width=0.49\textwidth]{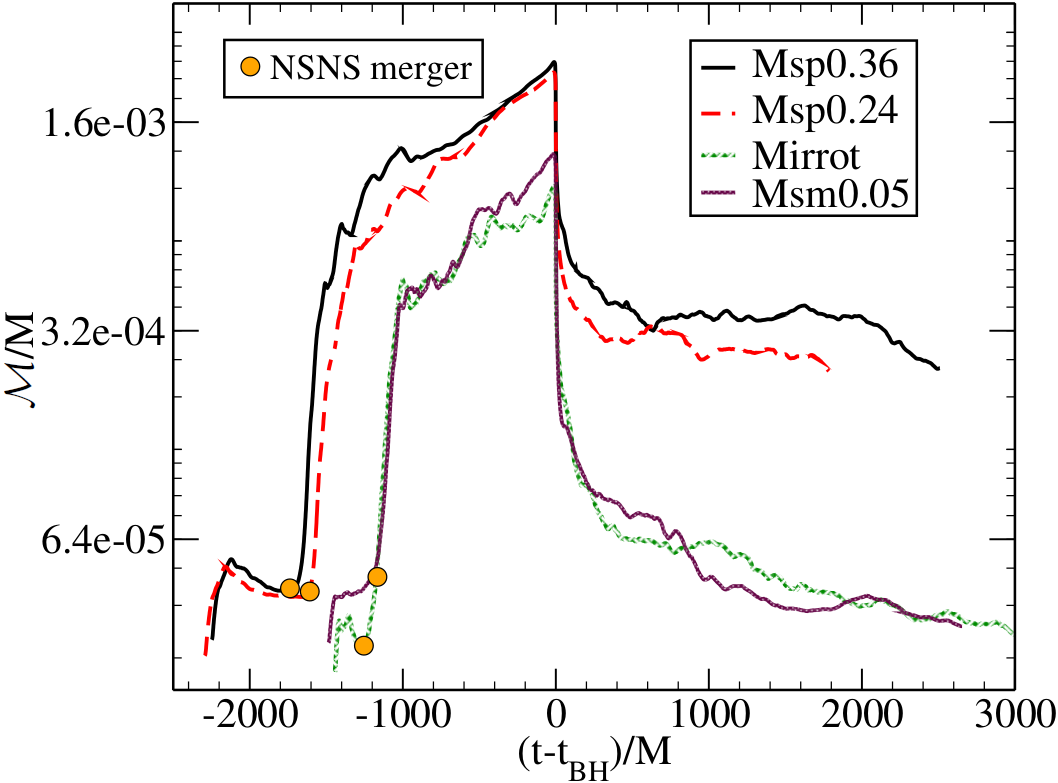}
  \caption{Total magnetic energy $\mathcal{M}$ normalized by the ADM mass
    $M=5.36\times 10^{54}(M_{\rm NS}/1.625M_{\odot})\rm erg$ versus time for
    cases listed in Table~\ref{table:summary_NSNSresults}. Dots indicate the
    NSNS merger time $t_{\rm mer}$. The coordinate time has been shifted to
    the BH formation time $t_{\rm BH}$.
    \label{fig:EM_outside}}
\end{figure}

Regardless, the angular momentum of the new-born magnetized HMNS,
dissipation of angular momentum by gravitational radiation, along with
transport angular momentum due to nonaxisymmetric torques, magnetic
winding and magnetic viscosity due to the MRI, cause the contraction
of the inner stellar region and the expansion of the external
layers. Eventually, the stellar inner region becomes a nearly
uniformly rotating massive core immersed in a Keplearian disk~(see
bottom panel in~Fig.~\ref{fig:Omega_plot}).  By $t-t_{\rm mer}\lesssim
1400M\sim 20.5(M_{\rm NS}/1.625M_\odot)\rm ms$ after merger the HMNS
collapses (see~Table~\ref{table:summary_NSNSresults}) as first
demonstrated in~\cite{Duez:2004nf}.  We find that the larger the spin
the smaller the mass of the remnant BH and, therefore, the heavier
the disk (see~Fig.~\ref{fig:M0_outside}). However, in all cases the 
BH dimensionless spin is $a/M_{BH}\simeq 0.78$
(see~Table~\ref{table:summary_NSNSresults}). 
The independence of the nascent BH spin on the
  initial NS spin may be an EOS-independent outcome. We plan to
  investigate this further in future work. Magnetic winding and
MRI will transport angular momentum as long as the matter is
differentially rotating. So, the HMNS will be driven into a massive
central core + disk configuration in a timescale that depends only on
how much angular momentum needs to be extracted from the inner stellar
region and deposited in the outermost layers (see Eqs.~\ref{eq:tau_w}
and \ref{eq:tau_mri}).  Note that in the high resolution case reported
in~\cite{Ruiz:2016rai} the remnant BH has a spin $a/M_{BH}\simeq
0.74$, though that case corresponds to the irrotational case in
Table~\ref{table:NSNS_ID} with a poloidal magnetic field confined to
the NS interior. High resolutions are therefore required to accurately
determine the final spin of the BH remnant.
Notice that in cases Msm0.05 and Mirrot, the HMNS collapses to a BH later
than its hydrodynamic counterpart (see Table~\ref{table:summary_NSNSresults}):
due to magnetic turbulence, kinetic energy is dissipated through small scale
shocks which heat up the star, thus increasing the thermal
pressure support compared to the scenario without magnetic fields.

%
\begin{figure}
  \centering
  \includegraphics[width=0.49\textwidth]{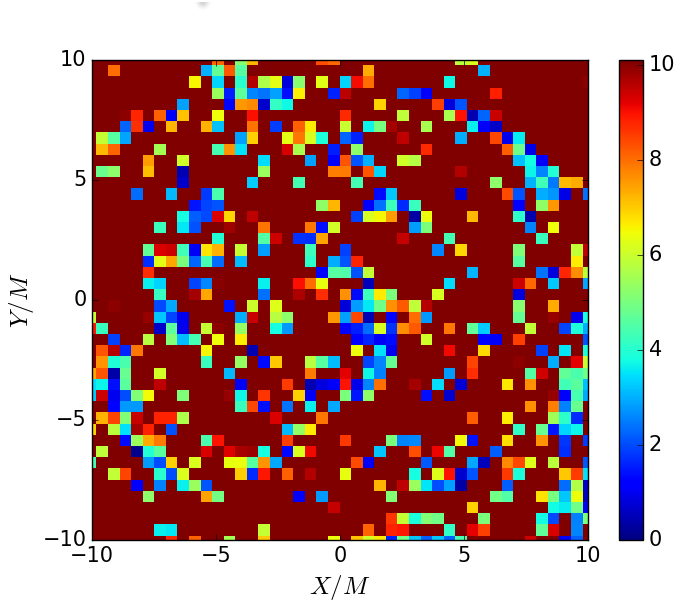}
  \includegraphics[width=0.48\textwidth]{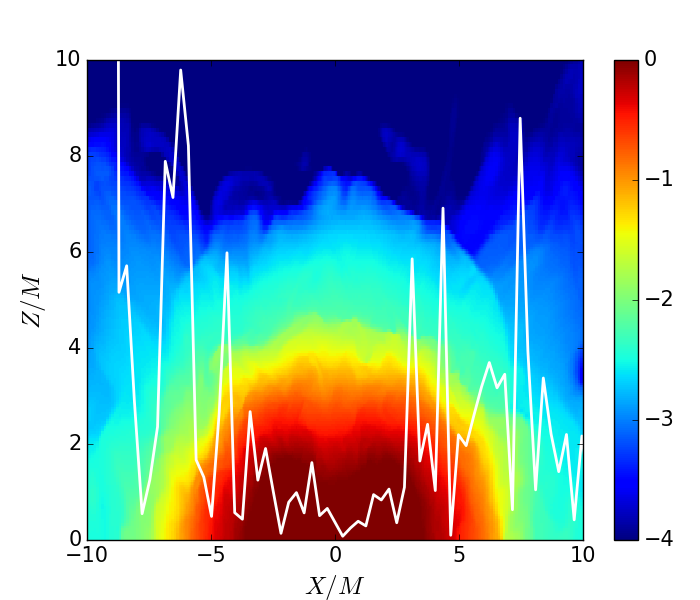}
  \caption{
    Contours of the quality factor $Q=\lambda_{\rm MRI}/dx$ on the equatorial plane (top panel),
    and rest-mass density, {{normalized to the initial maximum
        value $\rho_{0,\text{max}}\simeq 10^{14.4}(1.625\,M_\odot/M_{\rm NS})^2\text{g/cm}^3$}},
    and $\lambda_{\rm MRI}$ ({{white line}}) on the meridional plane
    (bottom panel) at $t-t_{\rm mer}
    \sim 400M\sim 6 (M_{\rm NS}/1.625 M_\odot)\rm ms$ for case Msp0.36.  We resolve the fastest growing
    MRI mode by $\gtrsim 10$ grid points over a large part of the HMNS. For most part
    $\lambda_{\rm MRI}$ fits within star. Other cases show similar behavior.
    \label{mri_alig}}
\end{figure}
%

\begin{figure}
  \centering
  \includegraphics[width=0.49\textwidth]{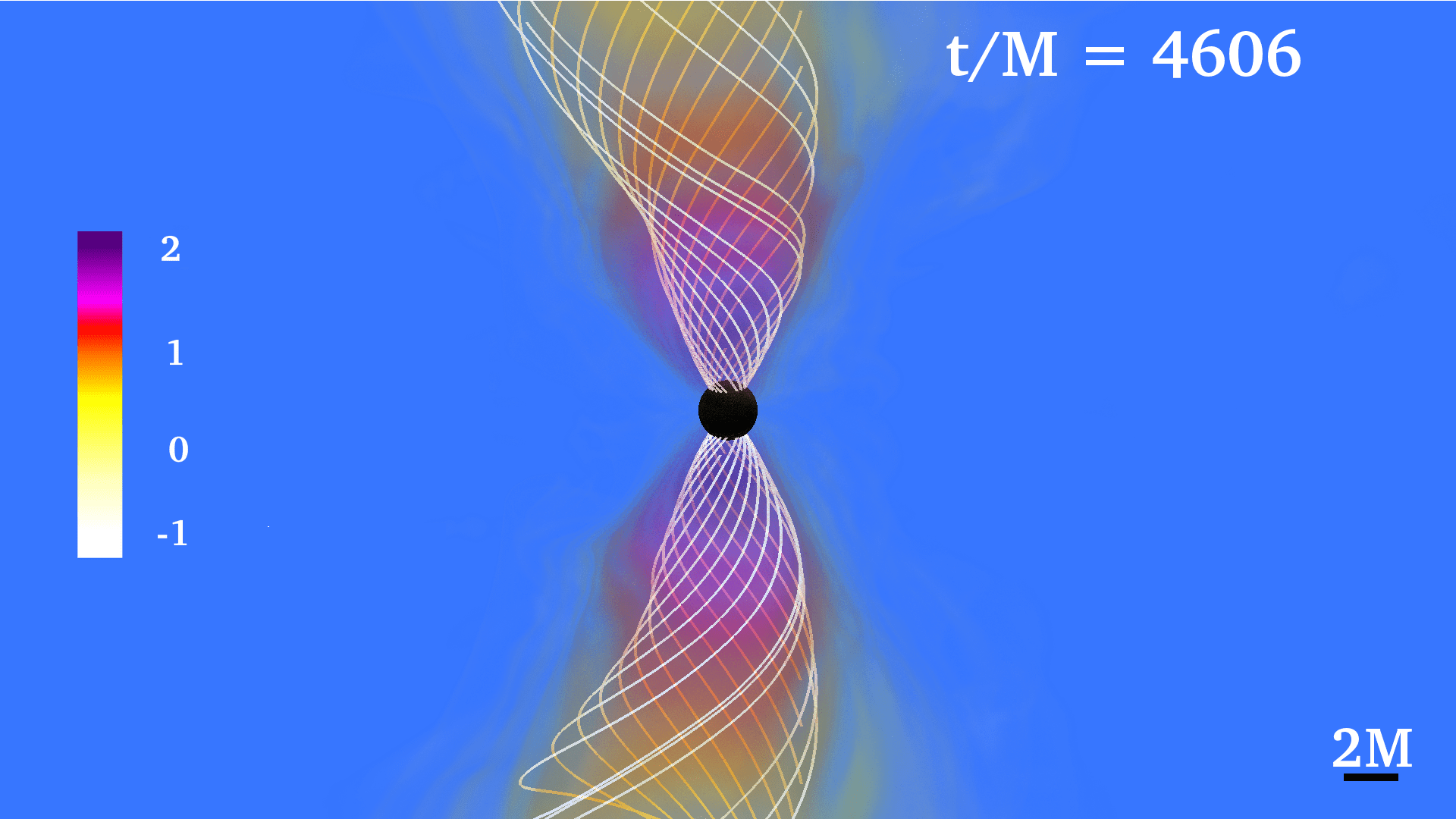}
  \caption{Volume rendering of the ratio~$b^2/2\rho_0$ (log scale) at $t-t_{\rm BH}\approx
    3400M\sim 50(M_{\rm NS}/1.625 M_\odot)\rm ms$ for the  Mirrot case, though similar behavior is
    observed in all magnetized cases.  Magnetic field lines (white lines) are plotted inside
    regions where~$b^2/2\rho_0 \gtrsim 10^{-2}$ (funnel boundary). Magnetically-dominated regions
    ($b^2/2\rho_0 \gtrsim 1$) extend to heights $\gtrsim 20M\sim 20\,r_{\rm BH}$ above
    the BH (black sphere). Here $r_{\rm BH}=2.2(M_{\rm NS}/1.625 M_\odot)\rm km$. 
    \label{fig:b2rho}}
\end{figure}

During HMNS collapse, the inner layers of the star, which contain most of the magnetic energy,
are promptly accreted into the BH. The magnetic energy $\mathcal{M}$ then plummets in only $t-t_{\rm BH}\sim
50{\rm M} \sim 0.7(M_{\rm NS}/1.625M_\odot)\rm ms$ following collapse, and then slightly decreases
thereafter as the accretion proceeds (see~Fig.~\ref{fig:EM_outside}). As magnetic winding during the
lifetime of the HMNS allows the magnetic energy to reach equipartition levels~\cite{Kiuchi:2015sga},
the magnetic field does not grow in the disk following BH formation~\cite{Ruiz:2016rai}. After HMNS
collapse, we find that the rms value of the magnetic field in the disk is $\lesssim 10^{16}(1.625
M_\odot/M_{\rm NS})\rm G$ (see~Table~\ref{table:summary_NSNSresults}).

Although immediately after BH formation the atmosphere is a very
gas-loaded environment~\cite{jojb15}, the winding of the magnetic
field above the BH poles has been well underway even before 
collapse (see fourth row
in~Fig.~\ref{fig:NSNS_snap}). By~$t-t_{BH}\sim 2000M\sim 30(M_{\rm
  NS}/1.625M_\odot)\rm ms$ in cases Msm0.05 and Mirrot,
and~$t-t_{BH}\sim 1000M\sim 15(M_{\rm NS}/1.625M_\odot)\rm ms$ in the
aligned cases, the magnetic pressure above the BH poles balances the
ram pressure of the fall-back debris and the inflow stops. Fluid
velocities then start to turn around and point
outward. Simultaneously, the magnetic field is tightly wound into a
helical funnel (see bottom panels in Fig.~\ref{fig:NSNS_snap}).  As
the regions above the BH poles are cleaned out,
magnetically--dominated regions ($b^2/(2\rho_0)\gtrsim1$) in the
funnel gradually start to expand. Once $b^2/(2\rho_0)\gtrsim 10$, the
magnetic pressure above the BH poles is high enough to overcome the
ram pressure, and a magnetically sustained outflow emerges. In all
cases, we observe that at about $t\sim 850M-900M\sim 12.5 13(M_{\rm
  NS}/1.625M_\odot){\rm ms}-13(M_{\rm NS}/1.625M_\odot)\rm ms$ after
the fluid velocities change direction for the first time, the outflow
reaches heights $\geq 100M \sim 430(M_{\rm NS}/1.625M_\odot)\rm km$,
and the Lorentz factor inside the funnel is $\Gamma_L \sim 1.1-1.3$.
Thus, at $\sim3000M-4000M\sim 45(M_{\rm NS}/1.625 M_\odot)\rm
ms-60(M_{\rm NS}/1.625 M_\odot)\rm ms$ following the NSNS merger a
magnetically-driven and mildly relativistic outflow 
--\textit{an incipient jet}-- has been launched
(see~Table~\ref{table:summary_NSNSresults}). Note that the jet near the poles
is only mildly relativistic.  However, as it is shown in Fig.~\ref{fig:b2rho}
the ratio $b^2/(2\,\rho_0)$ above the BH poles is $\gtrsim 100$. The
maximum feasible Lorentz factor $\Gamma_L$ for Poynting-dominated jets
equals $b^2/(2\,\rho_0)$ \cite{Vlahakis2003}. So, matter in the funnel
of the incipient jet can be accelerated to $\Gamma_L\gtrsim 100$, as
required by sGRB models.

To determine the collimation of the jet, we estimate the funnel opening angle $\theta_{\rm jet}$
using $b^2/(2\rho_0)\sim 10^{-2}$ contour as the boundary of the funnel~\cite{Ruiz:2016rai}. In all cases,
we find that the funnel opening angle is~$\sim 25^\circ-30^\circ$ (see~Fig.~\ref{fig:b2rho}). 
%
\begin{figure}
  \centering
  \includegraphics[width=0.49\textwidth]{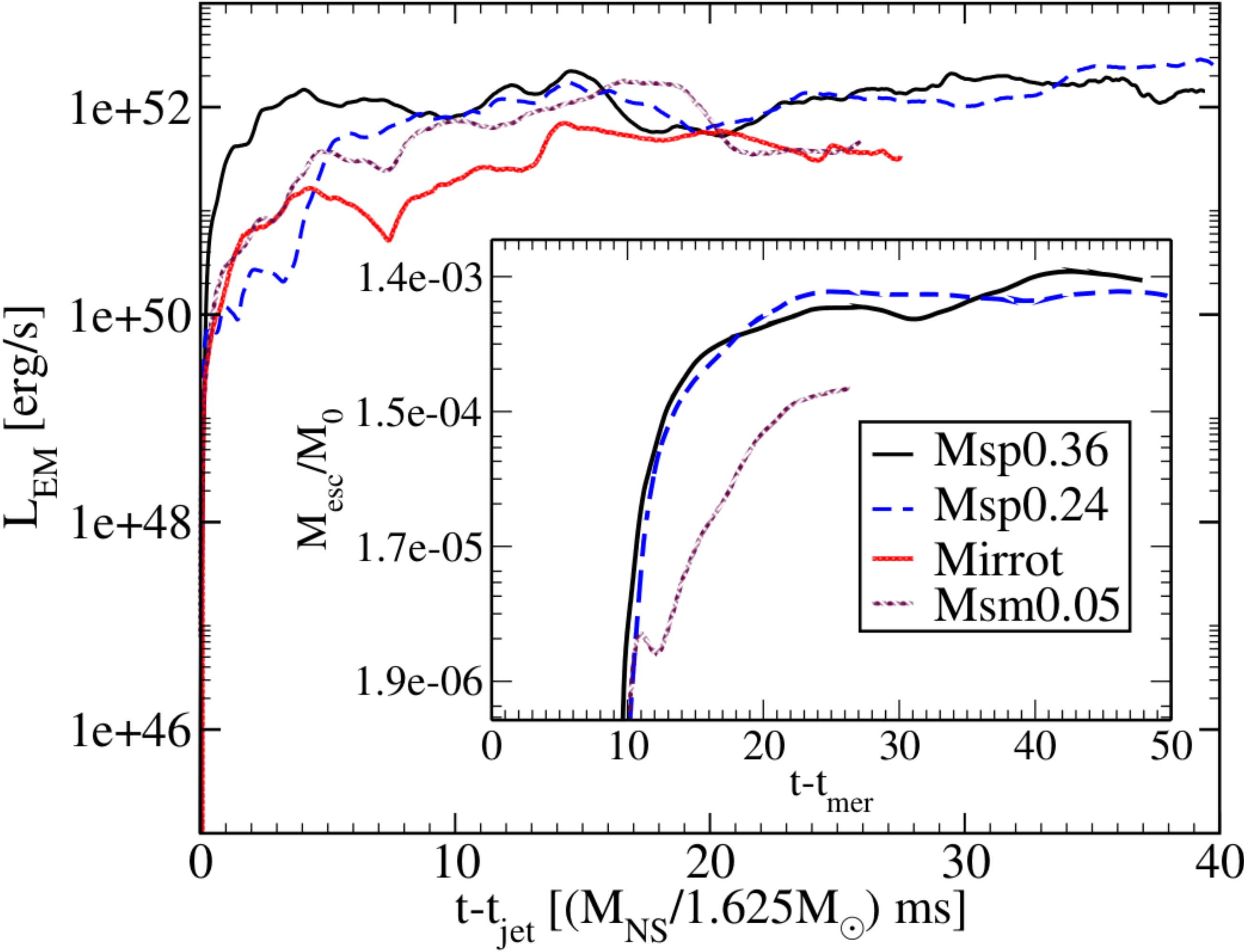}
  \caption{Outgoing EM (Poynting) luminosity for $t\geq t_{\rm jet}$ computed at a  coordinate sphere
    of  radius $r=120M\sim 530(M_{\rm NS}/1.625M_\odot)\rm km$ for the  magnetized cases listed in
    Table~\ref{table:summary_NSNSresults}. {{ The inset shows the rest-mass fraction of
    escaping matter during the last $\sim 2700M\sim 40
      (M_{\rm NS}/1.625M_\odot)\rm ms$ before the termination of our simulations.}}
    \label{fig:Poynting_plot}}
\end{figure}
Fig.~\ref{fig:M0_dot} shows the accretion disk history for all cases. In the aligned cases, the accretion
reaches a quasistationary state at about $t-t_{\rm BH}\sim 1000M\sim 15(M_{\rm NS}/1.625M_\odot)\rm ms$, while in the
other two cases the quasistationary state is reached at roughly  $t-t_{\rm BH}\sim 1350M\sim 20(M_{\rm NS}/
1.625M_\odot)\rm ms$.  Based on the accretion rate at $t-t_{\rm BH}\sim 2000 M\sim 30(M_{\rm NS}/1.625M_\odot)\rm
ms$ and the mass of the accretion disk (see~Fig.~\ref{fig:M0_dot}), we estimate  that the disk will be accreted
in $\Delta t\sim M_{\rm disk}/\dot{M}\gtrsim 98\rm ms$~(see~Table~\ref{table:summary_NSNSresults}), timescale
consistent with those of typical short-duration sGRBs~(see e.g.~\cite{Kann:2008zg}).

To verify if the BZ mechanism is operating in our simulations, we compare the outgoing Poynting luminosity
$L_{\rm EM}$ computed across a spherical surface of coordinate radius of $r_{\rm ext}\approx 120M\sim
530(M_{\rm NS}/1.625M_\odot)\rm km$ (see Sec.~\ref{subsec:diagnostics}) to the expected EM power generated
by the BZ mechanism given by~\cite{Thorne86} 
\begin{eqnarray}
  \label{eq:LBZ}
  L_{\rm BZ} &\sim&\\&& 10^{52} \left(\frac{a/M_{\rm BH}}{0.75}\right)^2\,
  \left(\frac{M_{\rm BH}}{2.8 M_\odot}\right)^2\left(\frac{|B|}{10^{16}\rm G}
  \right)^2\,\rm erg/s\,,
  \nonumber
\end{eqnarray}
As it is shown in Fig.~\ref{fig:Poynting_plot} the Poynting luminosity
is $L_{\rm EM}\sim 10^{51\pm 1}\rm erg/s$ (see
Table~\ref{table:summary_NSNSresults} for a time-averaged value over
the last $500M\sim 7.4(M_{\rm NS}/1.625 M_\odot)\rm ms$ before the
termination of our simulations when the jet is well--developed),
roughly consistent with the expected $L_{BZ}$ value in
Eq.~\ref{eq:LBZ}. In addition, we also compute the ratio of the
angular velocity of the magnetic field to the angular velocity of the
BH, $\Omega_F/\Omega_H$, on a meridional plane passing through the
apparent horizon centroid (see~e.g.~\cite{mg04}). We find that inside
the funnel $\Omega_F/\Omega_H$ ranges from $\sim 0.3-0.6$. As it has
been pointed out in~\cite{prs15,2004ApJ...611..977M}, deviations from
the expected value $\Omega_F/\Omega_H=0.5$ for an idealized monopole
field~\cite{Blandford1977} may be due to numerical artifacts or
deviations from strict force--free behavior. The above values along with the
tightly wound helical magnetic field above the BH poles suggest that
the BZ mechanism is likely operating in the BH + disk remnants.
We also compute the BZ power efficiency $\eta_{\rm eff}=L_{\rm EM}/\dot{M}$,
time-averaged over the last $t\sim 500M\sim 7.4(M_{\rm NS}/1.625M_\odot)\rm ms$
before the termination of our simulations. In all our models, we find that
$\eta_{\rm eff}\sim 0.3\%$ (see Table~\ref{table:summary_NSNSresults})~consistent
with  BH + disk GRMHD simulations for BHs with similar spinis (see Eq. 3
in~\cite{McKinney:2005zw}).
Note that the resulting luminosities and accretion rates are also
consistent with the ``universal model'' common to all BH + disk
systems formed following magnetized BHNS and NSNS mergers and
magnetorotational stellar collapse~\cite{Shapiro:2017cny}.

As it has been pointed out in~\cite{Metzger:2011bv}, matter ejection
in NSNS mergers $\gtrsim 10^{-3}M_\odot$ are required for detectable
kilonovae. {{
The inset in Fig. 11 shows the rest-mass fraction of escaping matter
during the last $\sim 2700M\sim 40 (M_{\rm NS}/1.625M_\odot)\rm ms$
out to our outer boundaries ($\sim 1100(M_{\rm NS}/1.625M_\odot)\rm km$)
before the termination of our simulations.  Our calculation does not
account for the ejected material that has left the numerical
domain by that time. Hence, our reported values indicate a {\it lower
limit} on the ejected material.}}
In our two aligned cases we find that the rest-mass fraction $M_{\rm esc}$ of the escaping
mass is $\gtrsim 10^{-2.3} (M_{\rm  NS}/1.625M_\odot)M_\odot$ (see inset in
Fig.~\ref{fig:Poynting_plot}).  So, in principle, the radioactive
decay of the above ejecta will power a light curve with a luminosity
of $\sim 10^{42}\rm erg/s$ and, therefore, could be detected by
current telescopes, as well as the LSST~\cite{Shibata:2017xdx,Rosswog}.
%
\section{Conclusions}
\label{sec:conclusion}

The likely assumption that the progenitor of GW170817 was a merging
NSNS system, along with the multiple counterpart radiation
observations across the EM spectrum, allows us to impose constraints on
the maximum mass of a nonrotating
star~\cite{Margalit:2017dij,Shibata:2017xdx,
  Ruiz:2017due,Rezzolla:2017aly}, on the radius of the
NS~\cite{TheLIGOScientific:2017qsa,Bauswein:2017vtn,Radice:2017lry,Most:2018hfd,
  Abbott:2018exr,Raithel:2018ncd}, and the EOS (see
e.g.~\cite{Annala:2017llu,Paschalidis:2017qmb,De:2018uhw,Abbott:2018exr}
and references therein). However, an open question for GW170817
  remains the impact of the initial NS spins on the outcome
  of the merger. These could have a strong
  impact on the remnant disk, the final BH spin, the lifetime
  of the transient HMNS, the amount of ejected
  neutron rich matter that can power kilonovae and synthesize heavy
  elements, and the formation and lifetime of a magnetically-driven
  jet and the associated outgoing EM Poynting luminosity. To address
  these issues, here we initiate GRMHD simulations of different NSNS
  configurations undergoing merger and delayed collapse to a BH while
  accounting for the initial NS spin. The binaries consist
  of two identical stars, in quasicircular orbit, each with spins
  $\chi_{\rm NS} = -0.053,\,0,\,0.24$, or $0.36$. In this first
  exploratory work we model the initial stars with a $\Gamma=2$
  polytropic EOS. To determine the impact of the magnetically-driven
effects on the fate of a spinning NSNS remnant, we have also considered
unmagnetized evolution of the above NSNS configurations.

We found that following the NSNS merger, the redistribution of angular
momentum due mainly to magnetic braking by winding and magnetic
turbulence driven by the MRI in the bulk of the transient HMNS
remnant, along with the angular momentum dissipation due to
gravitational radiation, induce the formation of a massive, nearly
uniformly rotating inner core immersed in a Keplerian disk-like envelope.
Eventually the HMNS collapses to a BH with a final spin $a/M_{\rm BH}
\simeq 0.78$ almost independent of the initial NS spin. In our 
  unmagnetized cases with high aligned spins the merger product is a long-lived
  HMNS (in contrast to the aligned spin magnetized models) while the
  irrotational and antialigned spinning ones collapse to a BH in agreement 
  with the pure hydrodynamic simulations of \cite{Kastaun2013,Kastaun2015} which
  used constraint violating spinning initial conditions. Regarding the 
  unmagnetized binaries, because of our relative low-mass priors we could not 
  verify the argument made by \cite{Kastaun2013,Kastaun2015} about the increase 
  of the BH spin vis-\`a-vis the initial spin of the NSs, and further work is 
  needed towards that direction. On the other hand the existence of a magnetic 
  field  triggers delayed collapse 
  (within $15(M_{\rm NS}/1.625 M_\odot)\ {\rm ms}$) in the case of sufficiently 
  low-mass remnants.
  The final BH in our simulations is surrounded by a magnetized accretion
disk whose rest-mass depends strongly on the initial spin of the NSs.
Our numerical results indicate that the excess of angular momentum is
deposited in the exterior layers to form the accretion disk. Thus, the
larger the initial spin of NSs the heavier the disk. We anticipate
that the above behavior will remain substantially unchanged when
alternative EOSs are used to model the NSs. Magnetic winding and MRI
will transport angular momentum as long as the matter is
differentially rotating. The HMNS will be then driven into a massive
central core + disk configuration in a timescale that depends only on
how much angular momentum needs to be extracted from the inner stellar
region and deposited in the outermost layers before centrifugal
support is no longer adequate to support the star against
collapse~(see Eqs.~\ref{eq:tau_w} and \ref{eq:tau_mri}).

After $\Delta t\sim 3000M-4000 M \sim 45(M_{\rm NS}/1.625 M_\odot)\rm
ms-60(M_{\rm NS}/1.625 M_\odot)\rm ms$ following merger, a
magnetically-driven and sustained incipient jet is launched. The
lifetime of the jets [$\Delta t\sim 100(M_{\rm NS}/1.625M_\odot){\rm
    ms}-140(M_{\rm NS}/1.625M_\odot)\rm ms$] and their outgoing
Poynting luminosities [$L_{\rm EM}\sim 10^{51.5\pm 1}\rm erg/s$] are
consistent with short-duration
sGRBs~\cite{Bhat:2016odd,Lien:2016zny,Svinkin:2016fho}, as well as
with the BZ process for launching jets and their associated Poynting
luminosities. The low luminosity of GW170817 [$L\sim 10^{47}\rm
erg/s$] is best understood by recent calculations showing that the jet
is misaligned with our line of sight by
$20^\circ-30^\circ$~\cite{Monitor:2017mdv}.

In the unmagnetized cases,  we found that, by contrast with the HMNS in the antialigned and irrotational
configurations that undergo delayed collapse to a BH after about $\sim 13 (M_{\rm NS}/1.625M_\odot)\rm ms$
following merger, the HMNS in the aligned cases is driven to a quasiaxisymmetric configuration on a dynamical
timescale, and remains in quasistationary equilibrium until the termination of our simulations ($t-t_{\rm mer}
\gtrsim 50(M_{\rm NS}/1.625M_\odot){\rm ms}$). Angular momentum redistribution by internal torques and dissipation
due to gravitational radiation alone are, therefore, inefficient mechanisms to trigger the collapse of a
highly spinning HMNS.


\acknowledgements
We thank the Illinois Relativity REU team (G. Liu, K. Nelli, and M. Nguyen) for
assistance with some of the visualizations. This work has been supported in part by National Science
Foundation (NSF) Grant  PHY-1662211, and NASA Grant 80NSSC17K0070 at the University of
Illinois at Urbana-Champaign. This work made use of the Extreme Science and Engineering Discovery
Environment (XSEDE), which is supported by National Science Foundation grant number TG-MCA99S008. This
research is part of the Blue Waters sustained-petascale computing project, which is supported by the
National Science Foundation (awards OCI-0725070 and ACI-1238993) and the State of Illinois. Blue Waters
is a joint effort of the University of Illinois at Urbana-Champaign and its National Center
for Supercomputing Applications.

\bibliographystyle{apsrev4-1}        
\bibliography{references}            
\end{document}